\documentclass{aastex63}

\received{....}
\revised{....}
\accepted{....}
\submitjournal{ApJ}

\usepackage{natbib,hyperref,ifthen,graphicx}
\usepackage{epstopdf,amsmath,longtable}
\usepackage{txfonts,amssymb}
\usepackage{soul,color}
\usepackage{rotating}
\definecolor{v}{rgb}{0.6, 0.2, 0.8} 
\defcitealias{Rodrigo2020}{Paper\,II}

\bibpunct{(}{)}{;}{a}{}{,}

\newcommand\beq{\begin{equation}}
\newcommand\eeq{\end{equation}}
\newcommand\beqn{\begin{eqnarray}}
\newcommand\eeqn{\end{eqnarray}}

\newcommand\kms{km s$^{-1}$}

\newcommand\sloanr{$r^\prime$}
\newcommand\sloani{$i^\prime$}
\newcommand\sloang{$g^\prime$}

\defcitealias{Gioia:1998}{G98}

\shorttitle{Dissecting the Strong Lensing Galaxy Cluster MS0440.5$+$0204}
\shortauthors{Verdugo et al.}
\begin{document}



\title{DISSECTING THE  STRONG LENSING GALAXY CLUSTER MS\,0440.5+0204\\
I. THE MASS DENSITY PROFILE\\}


\author{Tom\'as Verdugo}
\affiliation{Observatorio Astron\'omico  Nacional, Instituto de Astronom\'ia, Universidad Nacional Aut\'onoma de M\'exico, Ensenada, B.C., M\'exico; tomasv\makeatletter@astro.unam.mx}

\author{Eleazar R. Carrasco}
\affiliation{Gemini Observatory/AURA, Southern Operations Center, Casilla 603, La Serena, Chile.}

\author{Gael Fo{\"e}x}
\affiliation{Instituto de F\'{\i}sica y Astronom\'ia, Facultad de Ciencias, Universidad de Valpara\'iso, Avda. Gran Breta\~na 1111, Valpara\'iso, Chile.}

\author{Ver\'onica Motta}
\affiliation{Instituto de F\'{\i}sica y Astronom\'ia, Facultad de Ciencias, Universidad de Valpara\'iso, Avda. Gran Breta\~na 1111, Valpara\'iso, Chile.}

\author{Percy L. Gomez}
\affiliation{W. M. Keck Observatory, 65-1120 Mamalahoa Highway, Kamuela, HI, USA 96743.}

\author{Marceau Limousin}
\affiliation{Aix Marseille Univ, CNRS, CNES, LAM, Marseille, France.}

\author{Juan Maga\~na}
\affiliation{Instituto de Astrof\'isica, Pontificia Universidad Cat\'olica de Chile, Av. Vicu\~na Mackenna, 4860, Santiago, Chile.}

\author{Jos\'e A. de Diego}
\affiliation{Instituto de Astronom\'ia, Universidad Nacional Aut\'onoma de M\'exico, Avenida Universidad 3000, Ciudad Universitaria, C.P. 04510, Ciudad de M\'exico, M\'exico;}
\affiliation{Instituto de Astrof\'isica de Canarias (IAC), E-38200 La Laguna Tenerife Spain}


%


\begin{abstract}

We present a parametric strong lensing modeling of the galaxy cluster MS\,0440.5+0204 (located at $z$ = 0.19).  We have performed a strong lensing mass reconstruction of the cluster using three different models. The first model uses the image positions of four multiple imaged systems (providing 26 constraints). The second one combines  strong lensing constraints with dynamical information (velocity dispersion) of the cluster. The third one uses the mass calculated from weak lensing as an additional constraint. Our three models reproduce equally well the image positions of the arcs, with a root-mean-square image equal to $\approx$0.5$\arcsec$. However, in the third model, the inclusion of the velocity dispersion and the weak-lensing mass allows us to obtain better constraints in the scale radius and the line-of-sight velocity dispersion of the mass profile. For this model, we obtain $r_s$ = 132$^{+30}_{-32}$ kpc, $\sigma_s$ =   1203$^{+46}_{-47}$ km s$^{-1}$,  M$_{200}$ = 3.1$^{+0.6}_{-0.6}$ $\times10^{14}$\,M$_{\odot}$, and a high concentration, $c_{200}$ = 9.9$^{+2.2}_{-1.4}$. Finally, we used our derived mass profile  to calculate the mass up to 1.5 Mpc. We compare it with X-ray estimates previously reported, finding a good agreement.
\end{abstract}





\section{Introduction}

Studying  mass density profiles in galaxy clusters offers the opportunity to probe the formation and evolution of structures in the Universe and to probe different cosmological models \citep[see][]{Kravtsov2012}. In particular, three methods commonly used to study the mass in galaxy clusters are: 1.- strong and weak gravitational lensing \citep[e.g.,][and references therein]{Kneib2011,Hoekstra2013}, 2.- X-ray measurements \citep[see][and references therein]{Ettori2013}, and 3.- dynamical analysis from the velocity dispersion of the cluster members \citep[e.g.,][and references therein]{gir1998,Biviano2006,Old2015,Wojtak2018}. In fact, several authors have combined two or more of these probes to recover a more robust mass distribution in galaxy clusters \citep[e.g.,][]{kne03,Bradac2005,Newman2009,Morandi2012,Limousin2013,Verdugo2011,Verdugo2016,Siegel2018}. Indeed, even in the Hubble Frontier Field era, the study of the mass distribution in clusters requires data to probe the gravitational field beyond the nuclei responsible for strong lensing \citep{Limousin2016}.

This is the first paper (Paper\,I) in a series of two which aim to present a comprehensive and combined analysis of  \object{MS\,0440.5+0204}, performing lensing modeling (strong and weak), dynamical analysis (through spectroscopy of the galaxy members), and X-ray data. In \citet[][hereinafter \citetalias{Rodrigo2020}]{Rodrigo2020} we will show the detailed dynamical analysis of the cluster based in the redshift of 93 confirmed member galaxies inside 0.4 $\times$ 0.2 deg$^{2}$. In this first paper,  we present new  lensing modeles for  \object{MS\,0440.5+0204}, and compare the results with  the X-ray data available, and the mass previously reported by other authors \citep{Gioia:1998,Hicks2006,Shan2010}.

This galaxy cluster was selected for its regular morphology with not obvious substructures, and therefore easily 
  modeled by a single dark matter halo. The first  \object{MS\,0440.5+0204} model was presented in \citet{Gioia:1998}, hereafter  \citetalias{Gioia:1998}, 
when the only arc with measured redshift was A1 (See $\S$\,\ref{sec:data}).
Nevertheless, these  authors were able to put a limit
in the mass of the cluster and they calculated a
range of redshifts for the arcs. Using strong lensing \citet{wu00} estimated the
mass within the arcs ($9.0 \times10^{13}$
M$_{\sun}$), and compared it with the mass
calculated from the cluster X-ray luminosity,  finding that the former was smaller by a factor of two. This discrepancy was also reported by \citet{Shan2010} who analyzed a sample of 27 clusters including \object{MS\,0440.5+0204}. This cluster was also  explored  by different authors using weak lensing data \citep[e.g.,][]{hoekstra2012,Mahdavi2013}.  The study reported here is the first strong lensing analysis of  \object{MS\,0440.5+0204} since \citetalias{Gioia:1998},  using new spectroscopic redshifts for some of the multiple images as model constraints.

\begin{figure*}[!htp]
\epsscale{1.40}
\plotone{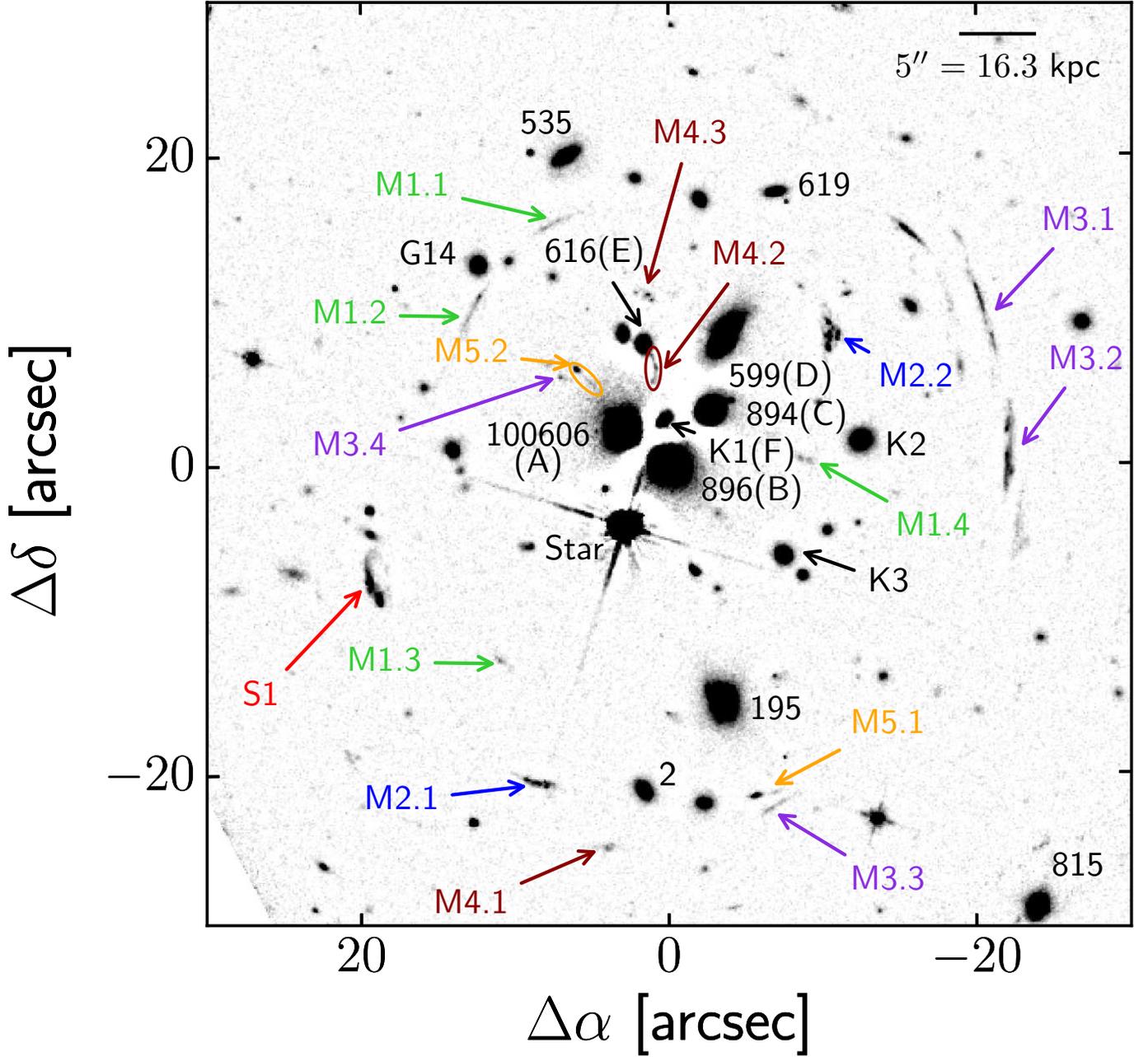}
\caption{  HST WFPC2 F702W image, with local median average subtracted of the central region ($60\arcsec \times60\arcsec$) of the MS0440 cluster. The center of the image coincides with the galaxy 896(B), located at $\alpha$  = $\mbox{4:43:09.8}$, $\delta$ = $\mbox{+2:10:18.2}$. The 14 galaxies considered in our models are identified along with the arcs systems; each color corresponds to a different arcs system. We highlight with ellipses the radial arcs reported  and discussed in \citetalias{Gioia:1998}.\label{fig1}}
\end{figure*}

In \citet{Verdugo2011} we combined strong lensing and dynamical information at large scale to estimate the scale radius of a Navarro-Frenk-White (NFW) mass profile \citep{nav96,nav97}, using the constraints as a prior in the strong lensing analysis. Here we aim  to extend the method of  \citet{Verdugo2011} by fitting strong lensing constraints with dynamical information, combining likelihoods, and adding the mass as an additional constraint. The paper is organized as follows: in the next section we describe the data used. We explain the methodology in section \S \ref{sec:method}.
In section \S \ref{sec:modeling} we 
present the different models studied in the paper and we discuss
the results in section \S \ref{sec:discussion}.
Finally, we present our conclusions in
section \S \ref{sec:conclusions}. Throughout this paper, we adopt a spatially flat
cosmological model dominated by cold dark matter and a
cosmological constant. We use $\Omega_{m}=0.3$,
$\Omega_{\Lambda}=0.7$ and H$_0$ = 70 km s$^{-1}$ Mpc$^{-1}$.
With these cosmological parameters, $1 \arcsec$ corresponds to 3.24 h$^{-1}$ kpc at the redshift of \object{MS\,0440.5+0204}.

\section{Observations}\label{sec:data}

\subsection{HST data}

The Hubble Space Telescope (HST) data have been obtained from the Multimission Archive
at Space Telescope (MAST). They consist of ten WFPC2
images\footnote{Proposal ID 5402} obtained in
the F702W filter with a total integration time of
22,200\,s.  The image reduction was performed using the IRAF/STSDAS\footnote{IRAF is distributed
  by the National Optical Astronomy Observatory, which is operated by
  the Association of Universities for Research in Astronomy, Inc.,
  under cooperative agreement with the National Science Foundation} package. First, a
warm-pixel rejection was applied to the images using the IRAF
task \emph{warmpix}. The cleaned images were then combined with
the task \emph{crrej} to remove cosmic-rays hits, and finally the image was aligned with WCS coordinates, i.e. north up, and east left.
Following \citetalias{Gioia:1998}, we subtracted an image composed of the median values of neighbors  (25 by 25 pixel box around each pixel) from the original.
This enhances the low-surface brightness structures that are lost in the envelope of the multiple-nucleus cD light distribution.
In Figure~\ref{fig1} we show the $60''\times60''$ field
obtained after the subtraction which is used in our
strong lensing analysis. It is important to note that The HST data used in the present paper are the same used by \citetalias{Gioia:1998}. Since the publication of that work, there have  been no new optical HST observation of \object{MS\,0440.5+0204}.

\begin{table*}
\caption{Properties of the arc systems. Columns 2 and 3
list the names used in this work and the original names given by
\citetalias{Gioia:1998} respectively. Column 6 list the measured spectroscopic
redshift (from \citetalias{Rodrigo2020}).  The last column shows in parenthesis the range of values predicted by
\citetalias{Gioia:1998}. }
\label{tbl-1} 
\centering 
\begin{tabular}{ccccccc}
\hline\hline 
\\
System & Arc &   Old name  &  \multicolumn{1}{c}{$\alpha$}
&  \multicolumn{1}{c}{$\delta$} &   $z_{spec}$  &    $z_{Gioia}$    \\
&   &  &  (J2000)
& (J2000)  &       &         \\
\\
\hline 
\\
\textrm{0} & $$ & $$  &  $$  &  $$  &  &   $0.53230^{\dagger}$ \\
$ $ & S1  & A1 & $4:43:11.10$ & $+02:10:10.19$ & $0.53223$  &   \\\\
I & $$ & $$  &  $$  &  $$  &   &  $[0.53,1.1]$   \\
$ $ & M1.1  &A8 & $4:43:10.36$ & $+02:10:33.48$ & 1.10125   &          \\
$ $ & M1.2  &A9 & $4:43:10.65$ & $+02:10:29.04$ & 1.10139  &       \\
$ $ & M1.3  &A12 & $4:43:10.55$ & $+02:10:05.50$ & 1.10176 &      \\
$ $ & M1.4  &A24 & $4:43:09.23$ & $+02:10:18.44$ &  1.10147  &      \\\\
II & $$ & $$  &  $$  &  $$  &    & $[0.60,1.6] $  \\
$ $ & M2.1  &A6 & $4:43:10.40$ & $+02:09:57.60$ &  0.95434   &           \\
$ $ & M2.2  &A5 & $4:43:09.13$ & $+02:10:26.49$ &  0.95426    &   \\\\
III & $$ & $$  &  $$  &  $$  &   &  $[0.59,\infty] $  \\
$ $  & M3.1  &A3 & $4:43:08.45$ & $+02:10:27.29$ &  $\ddagger$  &      \\
$ $  & M3.2  &A2 & $4:43:08.35$ & $+02:10:19.15$ &  $\ddagger$  &       \\
$ $  & M3.3  &A20 & $4:43:09.38$ & $+02:09:55.92$ &  ---  &     \\
$ $  & M3.4  & --- &  $4:43:10.29$  &  $+2:10:23.82$  & ---   &       \\\\
IV & $$ & $$  &  $$  &  $$  &  &  $[0.59,1.5] $  \\
$ $  & M4.1  &A18 & $4:43:10.09$ & $+02:09:53.41$ & --- &     \\
$ $  & M4.2  &A17 & $4:43:09.88$ & $+02:10:24.43$ & ---  &     \\
$ $  & M4.3  &A19 & $4:43:09.91$ & $+02:10:28.95$ & ---  &      \\\\
V & $$ & $$  &  $$  &  $$  &   &  $[0.59,1.5] $  \\
$ $  & M5.1  &A7 & $4:43:09.43$ & $+02:09:56.70$ & --- &    \\
$ $  & M5.2  &A16 & $4:43:10.20$ & $+02:10:23.98$ & --- &   \\
\\
\\
\hline 
\end{tabular}
\tablecomments{ ($\dagger$): Spectroscopic value reported by \citetalias{Gioia:1998}.  ($\ddagger$)There is a weak emission line present in the spectrum of both M3.1 and M3.2 at an observed wavelength of 7156\AA\, that may be  [OIII]$\lambda$2321 at $z$ = 2.0834.}
\end{table*}

\textit{Multiple images.-} In Figure~\ref{fig1} (see also Fig.~\ref{A00} in Appendix \ref{sec:ApendixA}) we identify the six arcs systems
used in our models. These systems were first reported and used by \citetalias{Gioia:1998}.
Systems consisting of a single arc
are denoted by S, while those showing multiple images are named by M. We have measured the redshifts of the arcs S1, 
M1.1, M1.2, M1.3, M1.4, M2.1, M2.2, M3.1 and M3.2 (see Sec.~\ref{GemData}, and Table~\ref{tbl-1}). These
measured redshifts confirms the proposed arcs systems associations used  by \citetalias{Gioia:1998} (even though
 we used these same systems, we have written our own nomenclature to point out how
the multiple images are related  to each other.)
No spectroscopic features were found in the spectra of M3.3, M4.1, M4.2, M4.3, M5.1 and M5.2.

Arc S1 appears to be a highly distorted image of a galaxy \citepalias{Gioia:1998},
with two substructures that are easily distinguished.
However, in spite of its appearance, it is likely  to be a single-image system since no counter images were found. Furthermore, our models are consistent with S1 being a single image (see Sec.~\ref{subsec:3models}). 
The images M2.1 and M2.2, at redshift $z$ = 0.9543, have multiple knots. The difference in the numbers of knots in each image
is because M2.2 is slightly elongated in the radial direction, splitting the inner part of the object in two different images, which are overlapped in M2.1.

Observing the image-pattern distribution of system M1, we note that it is similar (with a flip in parity) to that displayed in system M3. 
Additionally, M1.1 and M1.2 show a mirror-symmetry. Similar to the one in arcs M3.1 and M3.2, which are resolved into bright knots, with substructure that divided them in at least two parts. Thus, following \citetalias{Gioia:1998}, who assume that some arc systems
belong to the same source, we associate our measured redshifts of arcs M3.1 and M3.2 to its respective
counter-images, namely, M3.3, and M3.4. Note that system M3.4 it is not labeled in \citetalias{Gioia:1998}. This association is predicted by our models.

Near the cluster center there are two radial
structures: M4.2 and M5.2. These radial arcs were analyzed and discussed in \citetalias{Gioia:1998}, 
where the authors studied their geometry and calculated their magnitudes.
The importance of these arcs lies in their effect to constrain 
the mass profile. For an axially symmetric lens, the
Jacobian matrix of the lens mapping has two
eigenvalues: one relates the tangential critical
curve  to the total enclosed mass, and the other links 
the position of the radial arcs  to the derivative
of the mass \citep{sch92}.
These radial systems are key ingredients 
to characterize the profile in the inner part.
\citet{Mir95} showed that the combination of radial arcs
and their counter images provides powerful constraints
in the dark matter density profile as well as in the enclosed
mass. Because of the interest of the radial systems,
we have included them in our model even though we
do not have measured redshifts. Thus, we need to deduce whether  M4 and M5 
are images of one or two sources. Based in their
structure and shape, we infer 
that these images come from two different sources
(see Table~\ref{tbl-1}). The same reasoning was
used  to assume that M4.3  belongs to
system M4. Although we do not have measured redshifts for these systems, given their importance explained in the above lines, we keep them and set their redshifts as free parameters to be estimated during the modeling process.

\subsection{GEMINI data}\label{GemData}

The spectroscopic data used in this work will be  described in \citetalias{Rodrigo2020}.
We refer the interested reader to the forthcoming publication, \citetalias{Rodrigo2020}, for a detailed analysis. Here we offer a brief summary.
The images were obtained in 2011 (program ID: GS$-$2011B$-$Q$-$59) with the Gemini Multi-Object Spectrograph 
\citep[GMOS --][]{hook2004} at the Gemini South telescope in Chile. \object{MS\,0440.5+0204}
was imaged  in the  \sloang~($3 \times300$ sec exposures),  \sloanr~($3 \times200$ sec. exposures) and \sloani~filters 
 ($3 \times200$ sec. exposures). The images were observed during dark time and under photometric 
conditions, with  median values of seeing of 0\farcs51, 0\farcs50 and 0\farcs57 in \sloang~, \sloanr~ and 
\sloani~ filters, respectively. Galaxies for spectroscopic follow-up were selected using their magnitudes and colors. 
A total of  98 galaxies with \sloanr$\leq 22.5$ mag were distributed in four masks, two of them also included the
gravitational systems M1, M2, M3, M4, M5, and S1.

Multi-object observations (MOS) were carried out in 2013  (Program ID: GS$-$2012B$-$Q$-$53) under photometric conditions.
The spectra were acquired using the R400$+$ grating, 1\arcsec\  slit width, and $2\times2$ binning. Two masks were observed 
using a central wavelength of 6000\AA. The other two masks, were the faintest galaxies and the strongly lensed arcs were included,  the \textit{nod-and-shuffling} technique in \textit{band shuffling} mode was used with a  blocking filter OG515 and a central wavelength of 7500\AA. 
The total exposure time in each mask of 4$\times$1800\,s.  All spectra were reduced with the Gemini GMOS package pipeline, following the standard procedures for MOS and \textit{nod and shuffle} observations.

\textit{Arc redshifts.} While no spectroscopic features (at 1$\sigma$ level over the continuum) were found in the spectra of systems M4 and M5, we were able to determine the redshift of the lensed sources in the other four systems (M1, M2, M3, and S1, see Table~\ref{tbl-1}) The spectra of M1.1, M1.2, M1.3, and M1.4 show emission lines associated with [OII]$\lambda$3727,  at z = 1.10148 (average). In system M2, the spectra of arcs M2.1, and M2.2 show some emission lines ([OII]$\lambda$3727, H$_{\beta}\lambda$4863 and [OIII]$\lambda\lambda$4959,5007) placing the object at $z$ = 0.95430 (average). In the spectra of M3.1 and M3.2 a weak emission line is present, consistent with a redshift of  $z$ = 2.0834  for [OIII]$\lambda$2321.

\textit{Galaxy redshifts.}  The redshifts of the galaxies were determined using the programs
implemented in the IRAF \textit{RV} package. The galaxy spectra were separated in early- and
late-type populations. For galaxies with clear emission lines (late-type), the redshifts
were measured with the program  \textit{RVIDLINES}. For early-type galaxies, the redshifts
were measured by cross-correlating the spectra with high signal-to-noise templates using
the  \textit{FXCOR} program. We were able to determine the redshift of 99 galaxies.
We also include in our analysis 113 galaxy redshifts obtained by \citet{Yee1996} and 57 galaxy redshifts derived by \citetalias{Gioia:1998}, 
with some of the galaxies appearing in both catalogues. The final catalog contains redshift determination for 195 galaxies inside an
area of $\sim$ 0.4\degr $\times$ 0.2\degr around the \object{MS\,0440.5+0204} galaxy cluster. From those 195 galaxies, ten  (from \citetalias{Gioia:1998})  are located at the redshift of the cluster and were not covered by our observations. The average
redshift and the velocity dispersion of the cluster were calculated using the robust
bi-weight estimators of central location and scale \citep{beers1990} with the program
ROSTAT and  an iterative procedure that applies a 3-$\sigma$ clipping algorithm to
remove outliers.  We have obtained an average redshif of $\langle Z \rangle 
=0.19593^{+0.00033}_{-0.00031}$ ($58738^{+98}_{-92}$ \kms) and a line-of-sight (hereafter LOS)
velocity dispersion of $\sigma_{LOS} = 771^{+63}_{-71}$ \kms, with 93 member galaxies.

%

\begin{table*}
\caption{Best-fit model parameters.}
\begin{center}
\label{tbl-2} 
\begin{tabular}{lccccccccccccc}
\hline\hline 
\\
Parameter &
      \multicolumn{4}{c}{\textrm{M$_{\textrm{lens}}$}} &
      \multicolumn{4}{c}{\textrm{M$_{\textrm{lens-$\sigma_s$}}$}} &
      \multicolumn{4}{c}{\textrm{M$_{\textrm{lens-$\sigma_s$-mass}}$}} \\
     & Cluster &  896(B) &  599(D) & $L^{*}$ &  Cluster &  896(B) &  599(D) & $L^{*}$ &  Cluster &  896(B) &  599(D) & $L^{*}$  \\  \\
    \hline
    \\
X$^{\dagger}$ [\arcsec] & $4.1^{+0.6}_{-0.7}$ & -- & --  & --  &        $4.2^{+0.6}_{-0.7}$ & -- & -- & -- &  $3.6^{+0.9}_{-0.4}$ & -- & --  & --  \\ \\
Y$^{\dagger}$ [\arcsec] & $-2.0^{+0.8}_{-0.5}$&  --  &  --     &  --   &       $-1.4^{+0.3}_{-0.4}$ & -- & -- & -- &  $-1.9^{+0.7}_{-0.3}$ & -- &  -- & --  \\ \\
$\epsilon^{\dagger\dagger}$  & $0.25^{+0.04}_{-0.03}$ &  $0.69^{+0.07}_{-0.37}$   &    $0.56^{+0.08}_{-0.36}$   &  --   &        $0.26^{+0.03}_{-0.02}$ & $0.75^{+0.04}_{-0.38}$ & $0.36^{+0.11}_{-0.21}$ & -- &  $0.25^{+0.04}_{-0.02}$ & $0.78^{+0.01}_{-0.42}$ & $0.41^{+0.12}_{-0.25}$ & --  \\ \\
$\theta\,[^{\circ}]$  & $165.3^{+1.2}_{-0.7}$ &   $27^{+32}_{-17}$     &    $-37^{+19}_{-2}$   &  --  &        $165.8^{+0.8}_{-0.9}$  & $50^{+18}_{-32}$ & $-31^{+18}_{-6}$ & -- &  $166.3^{+0.6}_{-1.1}$& $15^{+35}_{-10}$ & $-38^{+20}_{-1}$ & --  \\ \\
$r_s$ [kpc] & $172^{+38}_{-62}$ &   -- &  --    & --   &        $137^{+32}_{-37}$  & -- & -- & -- &  $132^{+30}_{-32}$ & -- & -- & -- \\ \\
$\sigma_s$  [km s$^{-1}$] & $1248^{+61}_{-76}$ &   --  &  --  &  -- &         $1226^{+41}_{-62}$  & -- & -- & -- & $1203^{+46}_{-47}$& -- & -- & --  \\ \\
$c_{200}$  & $8.3^{+3.2}_{-1.0}$ &   --  &  --  &  -- &         $9.8^{+2.4}_{-1.4}$  & -- & -- & -- & $9.9^{+2.2}_{-1.4}$& -- & -- & --  \\ \\
$M_{200}$ [10$^{14}$M$_\odot$]   & $3.9^{+0.9}_{-1.2}$ &   --  &  --  &  -- &         $3.3^{+0.6}_{-0.8}$  & -- & -- & -- & $3.1^{+0.6}_{-0.6}$& -- & -- & --  \\ \\
$r_{\rm core}$ [kpc]   & -- & --  &   --  &  $[0.15]$      &  -- & -- & -- & $[0.15]$  & -- & -- & -- & $[0.15]$  \\ \\
$r_{\rm cut}$ [kpc]   & -- & $47^{+2}_{-22}$ &   $22^{+14}_{-13}$  &   $23^{+14}_{-13}$      &   --  &  $41^{+6}_{-20}$  & $17^{+7}_{-2}$ & $31^{+11}_{-16}$  & -- & $35^{+9}_{-16}$ & $23^{+4}_{-5}$  & $33^{+10}_{-21}$  \\ \\
$\sigma_0$ [km s$^{-1}$]  & -- & $325^{+53}_{-91}$   &   $499^{+1}_{-28}$  &   $124^{+38}_{-15}$       &   --  &  $314^{+59}_{-75}$  & $494^{+4}_{-32}$ & $115^{+35}_{-10}$  & -- & $279^{+89}_{-65}$  & $499^{+1}_{-26}$ &  $131^{+38}_{-22}$  \\ \\
\\
\hline 
\end{tabular}
\end{center}
\tablecomments{ ($\dagger$): The position in arcseconds relative to galaxy 896(B) located at $\alpha$  = $\mbox{4:43:09.8}$, $\delta$ = $\mbox{+2:10:18.2}$.($\dagger$$\dagger$):  The ellipticity is defined as $\epsilon$ = ($a^2$ $-$ $b^2$)/($a^2$ $+$ $b^2$), where $a$ and $b$ are the semi major and semi minor axis, respectively, of the elliptical shape. Some of the dark matter halo parameters of galaxies 896(B), and 599(D), were allowed to vary in the optimization procedure. Values in square brackets are not optimized.}
\end{table*}

\subsection{CFHT data}

MS\,0440.5+0204 was observed using MegaCam in the Canada- France-Hawaii-Telescope (CFHT). This cluster was part of the Canadian Cluster Comparison Project (CCCP), a comprehensive multi-wavelength study targeting 50 massive X-ray selected clusters of galaxies \citep[e.g.,][]{hoekstra2012}. The images were pre-processed using the {\tt Elixir} pipeline  at CHFT, and the resulting images were then calibrated  and combined into a single image using the {\tt Megapipe}  stacking pipeline \citep{gwyn08} at the Canadian Astronomy Data Center (CADC). The final \textit{g'} and \textit{r'} images have a total exposure  time of 2440 sec and 6060 sec with an average seeing of 0.89" and 0. 85", respectively. With  these data we constructed the catalogs used in Section~\ref{WL} to perform the weak lensing analysis.

\subsection{X-ray data}

\begin{figure}[htb!]
\epsscale{0.65}
\plotone{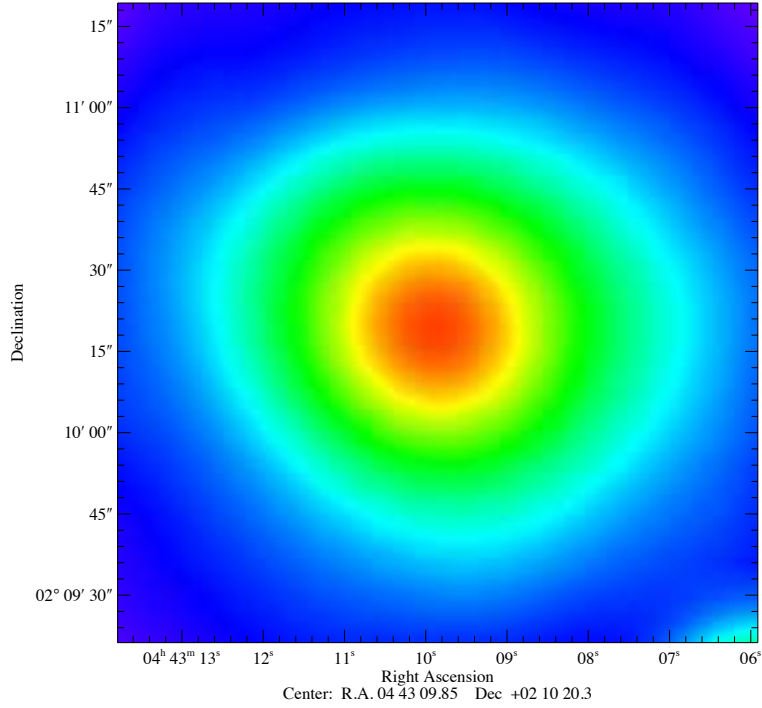}
\caption{Adaptively smoothed X-ray image of MS\,0440.5+0204 galaxy cluster in the 0.2-6.0 KeV energy range. The X-ray peak position coincides with galaxy 896(B). The image shows a slightly elongation, with a position angle in agreement with that obtained from strong lensing (see section~\ref{sec:discussion}).   \label{figXray}}
\end{figure}

The X-ray data of \object{MS\,0440.5+0204} was retrieved from the Chandra Data Archive. We processed the level 2 event files with the CIAO 4.11 software package. To map out the extended X-ray emission, we first removed point sources using the ‘wavedetect’ CIAO task. Next, we filtered the image using the 0.2-6 keV energy band. Finally, we adaptively smoothed the data with the ‘Csmooth’ task. The final image has been smoothed with kernels with minimum scale of 0.5 arcsec and maximum scale of 5 arcsec, with a minimum signal to noise of 3 arcsec per kernel. The smoothed Chandra image of the cluster in the 0.2-6 keV energy band is shown in Figure~\ref{figXray}.

The peak of the X-ray image coincides with the bright central galaxies, and the X-ray isophotes are elongated in the same direction as the mass contours obtained from the lensing analysis (see Section ~\ref{sec:discussion}). The final X-ray image has a higher spatial resolution than the one reported by \citep{Gioia:1998}. Note that  this is a new reduction to the same data used by \citet{Hicks2006} and \citep{Shan2010}. Notice that this X-ray data is used to provide morphological information, however the X-ray mass estimation supplied by the literature are compared to our lensing mass results in section~\ref{sec:discussion}.

\section{The method}\label{sec:method}

In this section we present a general review of the mass density profile used in our models, and derive some useful relations. We also define our different figure-of-merit-functions, and explain how they are constructed and computed in our models.

\subsection{Mass density profile}

We model the mass density profile of \object{MS\,0440.5+0204} as a NFW profile \citep{nav96,nav97}: 

\begin{equation}\label{eq:rho}
\rho(r) = \frac{\rho_{s}}{(r/r_s)(1+r/r_s)^{2}}.
\end{equation}

\noindent This profile, predicted in cosmological $N$-body simulations, is characterized by the  scale radius $r_s$,  that corresponds to the region where the logarithmic slope of the density equals the isothermal value, and the density $\rho_s$. The scale radius is related to the radius $r_{200}$ through the expression $c_{200}$ =  $r_{200}/r_s$, which is commonly called the concentration. The $r_{200}$ parameter is defined as the radius of a spherical volume inside of which the mean density is $200$ times the critical density, $\rho_{crit}$, at redshift $z$, i.e., $M_{200} = 200 \times (4\pi/3)r_{200}^{3} \rho_{crit}(z)$. By integrating Eq.~\ref{eq:rho}, it is straightforward to show that the mass contained within a radius $r$ of  the NFW halo is given by

\begin{equation}\label{eq:mass}
 \tilde{M}(r)  =   4\pi r_s^{3}\rho_s \left[\ln{(1+r/r_s}) - \frac{r/r_s}{1+r/r_s}\right].
\end{equation}

Similarly, but integrating along one axis, we obtain the surface mass density of the NFW profile

\begin{equation}\label{eq:SufaceMass}
\Sigma(\xi) = 2\rho_sr_sF(\xi),
\end{equation}

\noindent where $F(\xi)$ is a function \citep[e.g.,][]{gol02} of the dimensionless coordinate $\xi$,  the radius in the $XY$ plane in units of the scale radius,  $\xi$ = ($x/r_s$, $y/r_s$). Integrating the surface mass density, we obtain the  2D mass inside the radius $\xi$

\begin{equation}\label{eq:Mass2D}
 \tilde{m}(\xi) = 4\pi\rho_sr^3_sg(\xi),
\end{equation}

\noindent where  $g(\xi)$ is an expression that can be calculated analytically \citep[][]{bar04}.

The mass and the density profile are related to the velocity dispersion (at radius $r_s$) in the LOS through the expression \citep{Mamon2005,Mamon2013}

\begin{equation}\label{eq:sigmaNFW_pro}
 \tilde{\sigma_s}^{2}   = \frac{2G}{\Sigma(r_s)}\int_{r_s}^{\infty}  \mathcal{K}(r/r_s)  \tilde{M}(r)\rho(r) \frac{dr}{r},
\end{equation}

\noindent where $\mathcal{K}(r/r_s)$ = $\sqrt{1-(r_s/r)^2}$ for an isotropic velocity dispersion. Note that in this case, Eq.\,\ref{eq:sigmaNFW_pro} is equivalent to Eq.\,7 in \citet{Verdugo2011}.

 \subsection{The figure-of-merit-functions}

The $\chi^{2}$ has been previously used to quantify the goodness of fit of the lens model  \citep[e.g.,][]{Verdugo2007,Verdugo2011,jul07},  therefore, we summarize the method here.  Let us assume a model whose parameters are $\vec {\theta}$, with $N$ sources, and $n_i$ the number of multiple images for source $i$.  For every system $i$, we compute  the position in the image plane $x^j(\vec {\theta})$ of image $j$, using the lens equation. Therefore, the total $\chi^{2}$ from multiple image system $i$ is

\begin{equation}\label{eq:Chi2Lens}
\chi_{i}^{2} =     \sum_{j=1}^{n_i}
\frac{\left[ x_{obs}^j - x^j(\vec {\theta})    \right]^2}{\Delta_{ij}^{2}},
\end{equation}

\begin{figure}[htb!]
\epsscale{0.6}
\plotone{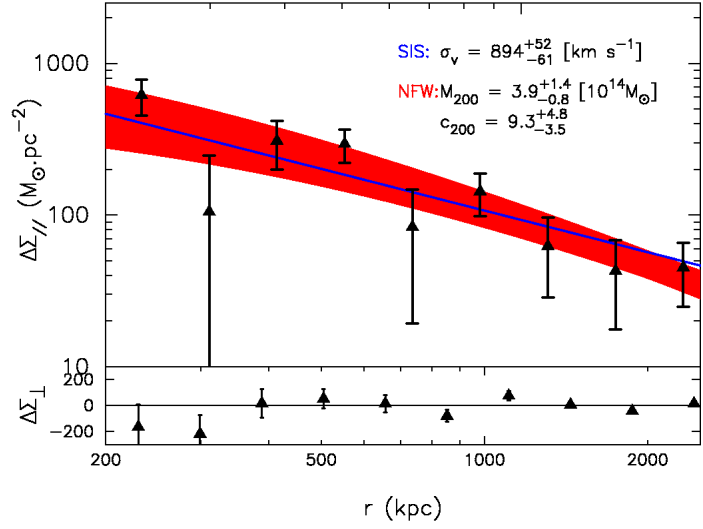}
\caption{Average density contrast $\Delta\Sigma$. The upper panel,  $\Delta\Sigma_{\parallel}$, is the profile using the tangential component. Over-plotted are the best-fit results for the two mass models, SIS (continuous blue line), and NFW (red-shaded area, encompassing the combined 1-$\sigma$ uncertainty on M$_{200}$ and $c_{200}$). The lower panel shows the profile obtained using the radial component of  the lensed galaxy's ellipticity,  $\Delta\Sigma_{\perp}$, and should be  equal to zero. \label{fig2}}
\end{figure}

\noindent were $\Delta_{ij}$ is the measured error in the position of image $j$, and $x_{obs}^j$ is the observed position. The $\chi_{i}^{2}$ from all image systems are added to form a $\chi^2_{\textrm{lens}}$.

We also define a $\chi^{2}_{\sigma_s}$ associated with the velocity dispersion as

\begin{equation}\label{eq:Chi2Sigma}
\chi^2_{\sigma_s} = \frac{\left[\sigma_{obs}-\tilde{\sigma_s}(\rho_s,r_s)\right]^2}{\Delta_s^{2}}
\end{equation}

\noindent where $\sigma_{obs}$ is the velocity dispersion obtained from the spectroscopic analysis \citepalias{Rodrigo2020},  $\Delta_s$  is the  error,   and $\tilde{\sigma_s}(\rho_s,r_s)$ is calculated through Eq.\,\ref{eq:sigmaNFW_pro}. Note that we are assuming that the velocity dispersion at $r_s$ is similar to the velocity dispersion measured at a greater radius.

Finally, the weak lensing analysis provide us with a 2D mass for the system measured at radius $\xi$, thus, we define

\begin{equation}\label{eq:Chi2X}
\chi^2_{\textrm{mass}} = \frac{\left[m_{WL}(\xi)-\tilde{m}(\rho_s,r_s,\xi)\right]^2}{\Delta_m^{2}}
\end{equation}

\noindent were $\Delta_m$ is the error in the mass, and $\tilde{m}(\rho_s,r_s,\xi)$ is calculated through expression \ref{eq:Mass2D}. Hence, with a likelihood  $\mathcal{L} \propto exp(-\chi^2/2)$, we define

\begin{equation}\label{eq:Chi1}
\chi^2_{\textrm{lens-$\sigma_s$}} =  \chi^2_{\textrm{\textrm{lens}}} + \chi^2_{\sigma_s},
\end{equation}

\noindent with $\chi^2_{\textrm{lens}}$,  and  $\chi^2_{\sigma_s}$ being the statistical parameters obtained from the fit. We also define

\begin{equation}\label{eq:Chi2}
\chi^2_{\textrm{lens-$\sigma_s$-mass}} =  \chi^2_{\textrm{\textrm{lens}}} + \chi^2_{\sigma_s} + \chi^2_{\textrm{mass}},
\end{equation}

\noindent where we have added the $\chi^2_{mass}$ related to the mass.

\begin{figure}[htb!]
\epsscale{0.6}
\plotone{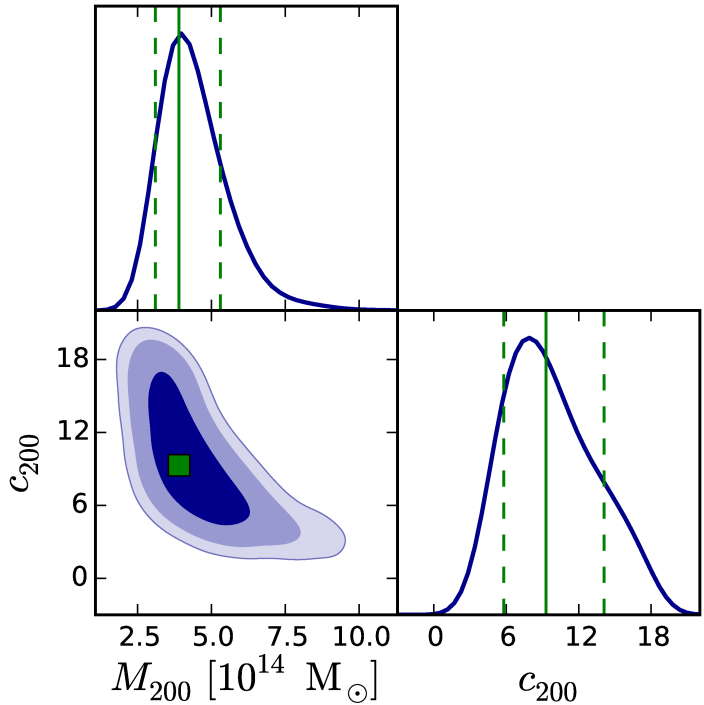}
\caption{ PDFs and contours of $c_{200}$ and M$_{200}$  parameters using weak lensing analysis. The three contours stand for the 68$\%$, 95$\%$, and 99$\%$ confidence levels. The values obtained for our best-fit model are marked by a green square and vertical lines in the 2D contour and in the 1D histograms respectively (the asymmetric errors are presented in Table 2).\label{fig3}}
\end{figure}

Note that we are not performing a simultaneous fit between data sets, for example, as those conducted by \citet{Verdugo2016} with dynamics and strong lensing, or the ones by \citet{new13} using weak lensing and strong lensing. Our approach is more simple, the idea is to construct a model that is consistent with other data sets that are calculated independently. In other words, the mass and velocity are added only as constraint to the modeling. This is an alternative when the modeler does not have access to the original data to perform a simultaneous fitting.

Finally, we  assume the same weight for strong lensing, dynamics, and weak lensing in the total likelihood, but when combining SL and weak lensing this could be different.  As discussed in \citet{Umetsu2015}, the contribution of strong lensing to the total fit could  bias the result if the image systems having similar configurations are not taken into account. However, this is not the case in the model of \object{MS\,0440.5+0204}, where the four systems are  different.

\section{Mass Modeling}\label{sec:modeling}

\subsection{Strong Lensing} \label{SL}

We operate our models with the last version of the
LENSTOOL\footnote{\tt This software is publicly available at:
http://projets.lam.fr/projects/lenstool/wiki} ray-tracing code,
which uses a Bayesian  Markov chain Monte Carlo (MCMC) method
\citep{jul07}. To model the dark matter component of this cluster, we consider first a single large scale clump, and then we add small-scale clumps as perturbations  associated with individual cluster galaxies.  A single clump was chosen given the 
elliptical pattern of the lensed arcs and the lack of a secondary luminous sub-clump. In addition, the single cluster model has sufficient accuracy (quantified by a small $\chi^2$) to reliably reproduce the image positions of the arcs.

 \textit{Large scale clump.-}  This component was modeled as a NFW mass density profile (see Eq.\,\ref{eq:rho}). 
 To take into account  the ellipticity in the lens modeling,
we consider the \emph{pseudo-elliptical} NFW proposed by \citet{gol02}.
This potential is characterized by six parameters: the position
$X,Y$; the ellipticity $\epsilon$; the position angle
$\theta$; the scale radius $r_s$ and the velocity dispersion
$\sigma_s$. In practice, all the parameters describing the
large scale dark matter clump are allowed to vary in the optimization procedure. In particular, the priors for the velocity and the scale radius are broad, and set as follow: 780  km s$^{-1}$ $\leq$ $\sigma_s$ $\leq$ 1600  km s$^{-1}$, and 50 kpc $\leq$ $r_s$ $\leq$ 400 kpc.

\textit{Small scale clumps.-} The cluster galaxy population is incorporated
into the lens model as pseudo-isothermal elliptical mass distribution potentials \citep[PIEMD,][]{lim05,eli07}. This mass distribution is parameterized by a  velocity dispersion, $\sigma_0$, related to the central density of the profile.
It has two characteristic radii, $r_{core}$ and $r_{cut}$, that define changes in its slope.  In the inner region, the profile has a core with  a central density $\rho_0$, then in a transition region ($r_{core}$ $< r <$ $r_{cut}$) it becomes isothermal, with $\rho$ $\sim$ $r^{-2}$. These parameters
are scaled as a function of luminosity:

\begin{equation}\label{eq:scale}
\begin{array}{l}
r_{core}=r_{core}^{*}(\frac{L}{L^{*}})^{1/2},\\ \\
 r_{cut}=r_{cut}^{*}(\frac{L}{L^{*}})^{1/2},\\ \\
 \sigma_0 = \sigma_0^{*}(\frac{L}{L^{*}})^{1/4}.
\end{array}
\end{equation}

\begin{figure}[h!]\begin{center}
\includegraphics[scale=0.72]{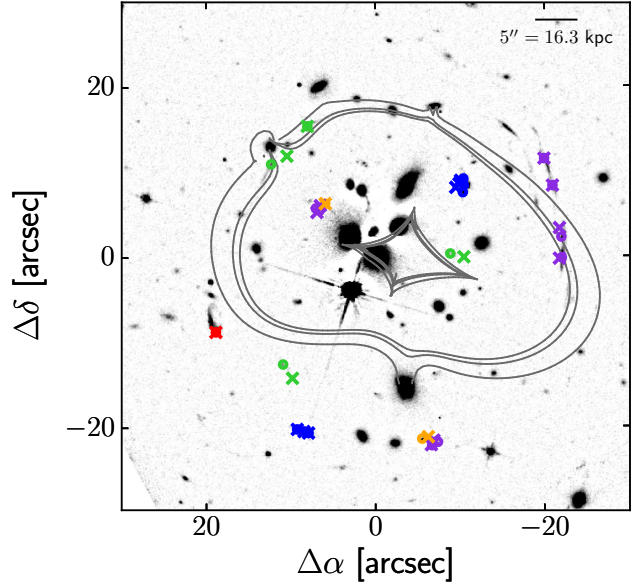}
\caption{Critical and caustic lines for model M$_{\textrm{lens}}$ (for a source at $z$ = 0.95, $z$ = 1.10, and $z$ =  2.08, from inner to outer radii respectively). The circles show the positions of the images (input data for the model), and the crosses the predicted positions of the lensed images. We follow Fig.1 code-colors for each family of lensed images.  \label{critics}}
\end{center} 
\end{figure}

\noindent The scaling relation for $\sigma_0$
assumes that mass traces light, and its origin
resides in the Faber-Jackson relation
\citep{fab76}, that has been reliable for
describing early-type cluster galaxies \citep[e.g.][]{wuy04,fri05}. These scaling relations are common in lens 
modeling techniques\citep[e.g.,][]{lim07b}.

The 13 member galaxies in the central part of the cluster (see Figure\,\ref{fig1}) 
are early-type and, therefore, they satisfy the above
scaling relations (Eq. \ref{eq:scale}). 
Then, for a given $L^{*}$ luminosity, we will search for those values of $\sigma_0^{*}$ and $r_{cut}^{*}$
that yield the best fit, while $r_{core}^{*}$
is fixed at 0.15\,kpc  \citep[the value is arbitrary, see discussion in][]{lim07b}. These two parameters, $\sigma_0^{*}$ and $r_{cut}^{*}$, describing
the cluster galaxy population, add two more free parameters
in the optimization procedure. For
a galaxy with $r'$ magnitude equal to
17.70 (galaxy 100606(A), see \citetalias{Rodrigo2020}), we
set the following limits:
150 km s$^{-1}$ $<\sigma_0^{*}<$ 300 km s$^{-1}$, and
5 kpc $<r_{cut}^{*}<$ 50 kpc.
These limits in the parameters are motivated
by galaxy-galaxy lensing studies in clusters
\citep{nat98, nat02a, nat02b,lim07a},
and numerical simulations  \citep{Limousin2009}. The other parameters describing the galaxy
scale clumps are set as follows: the
center of the dark matter halo is assumed
to be the same as for the luminous component,
and the ellipticity and position angle of the mass
is assumed to be the same as those of the
light. The luminosity distribution of a given galaxy may not trace the
dark matter distribution in its halo, however there is evidence
that the projected mass and light distributions tend to be
aligned \citep{kee98}.

\begin{figure*}\begin{center}
\includegraphics[scale=0.35]{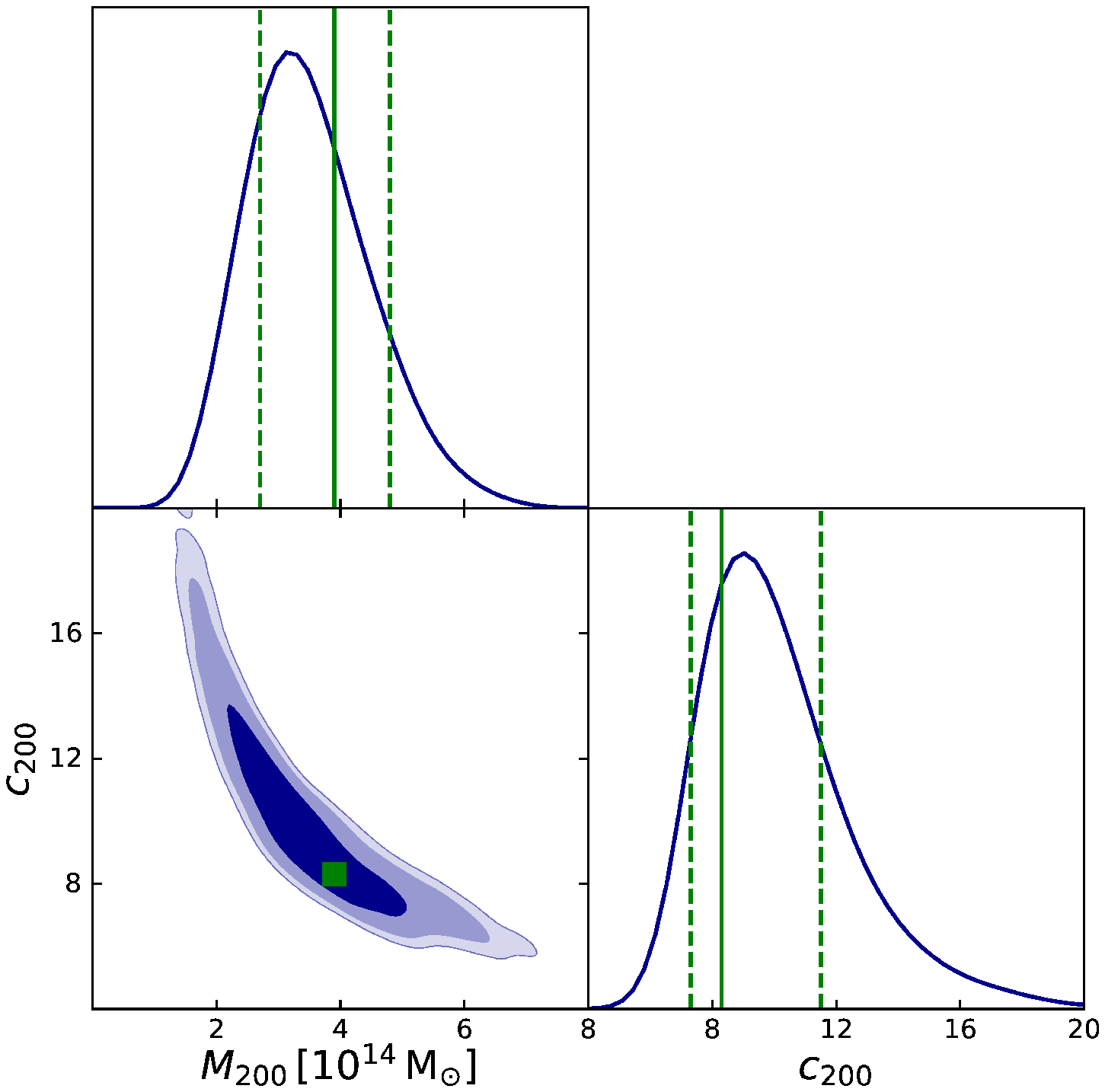}
\hspace{0.1cm}
\includegraphics[scale=0.35]{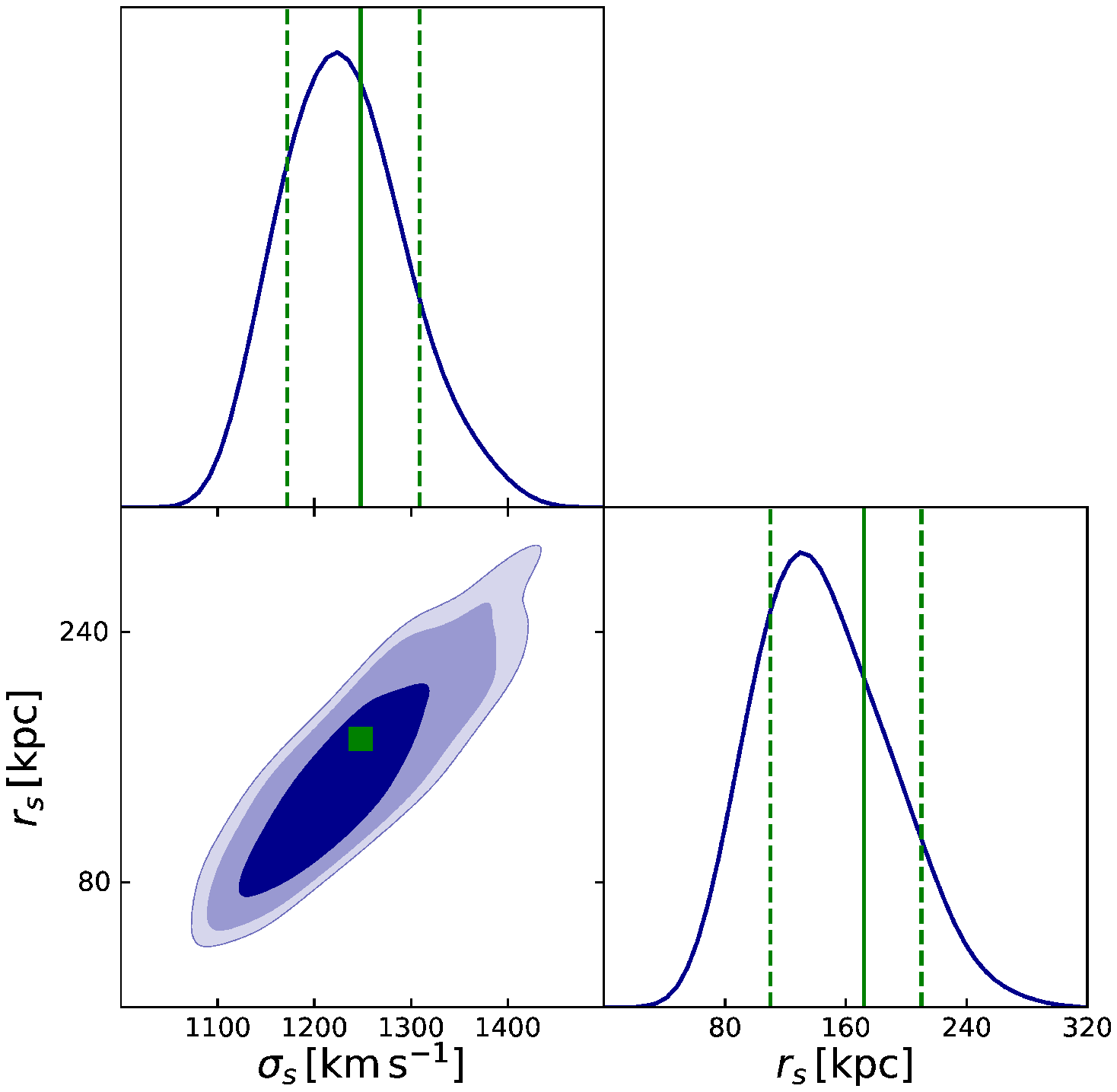}\\
 \vspace{0.1cm}
\includegraphics[scale=0.35]{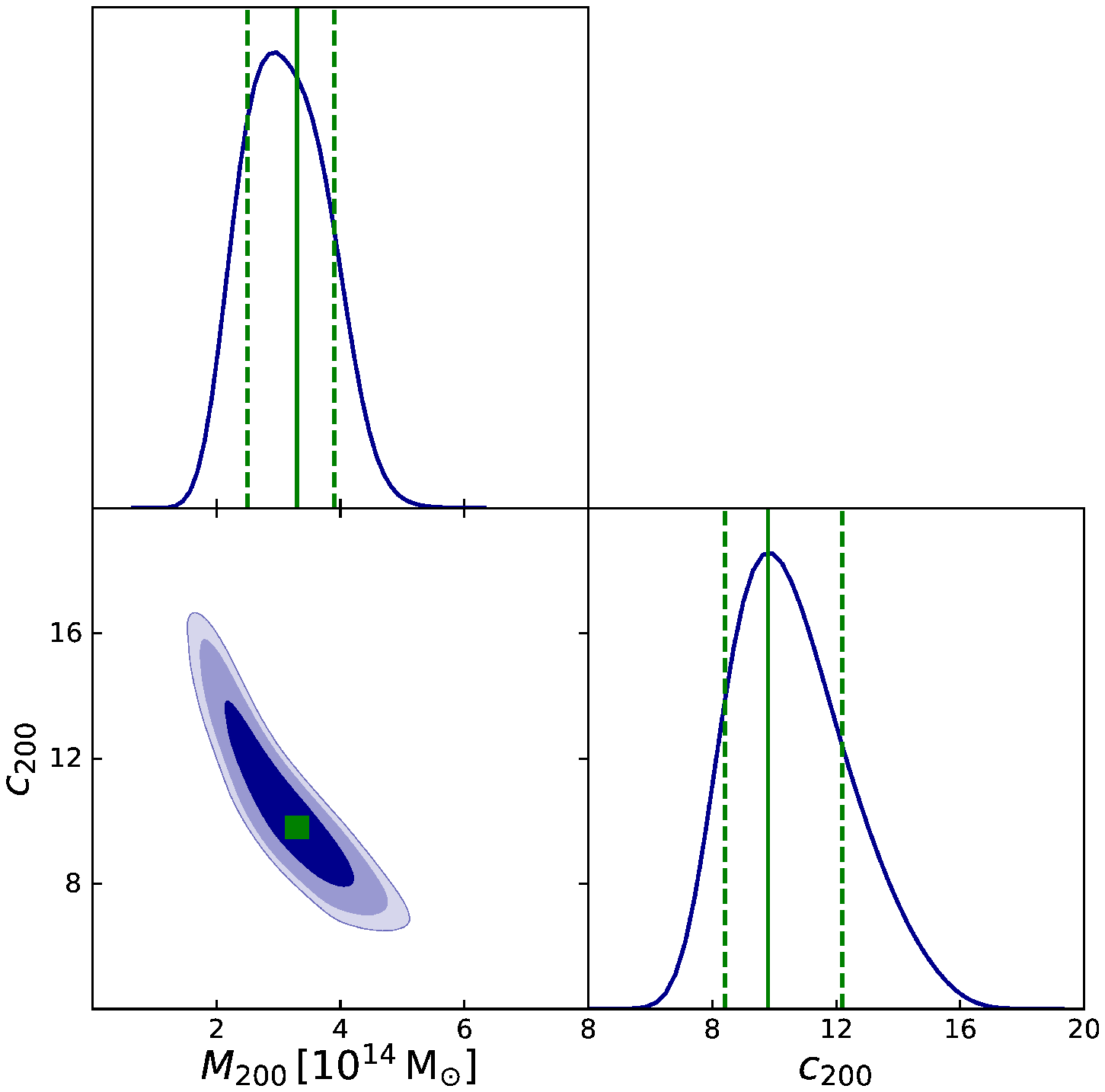}
\hspace{0.1cm}
\includegraphics[scale=0.35]{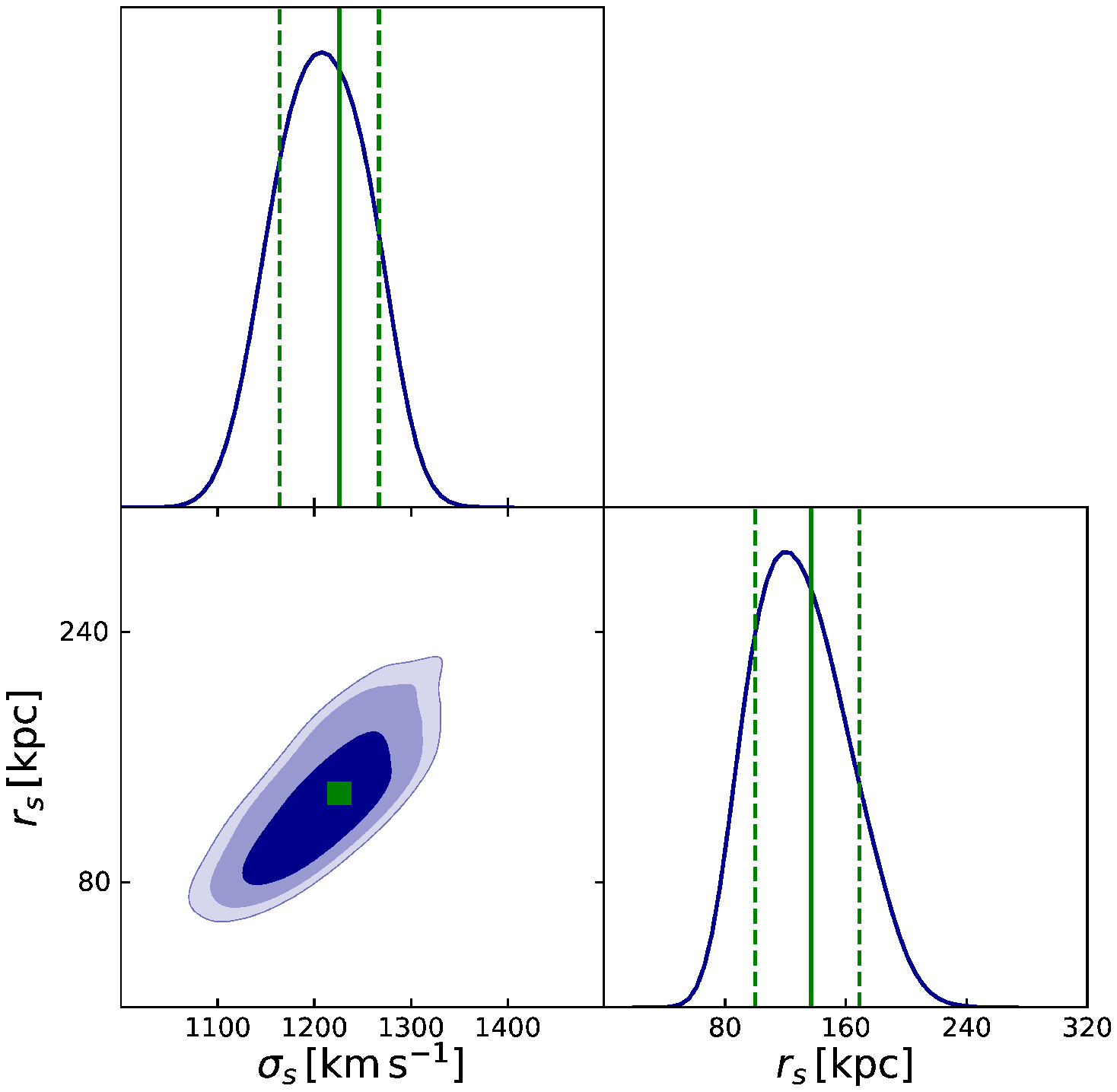}\\
 \vspace{0.1cm}
\includegraphics[scale=0.35]{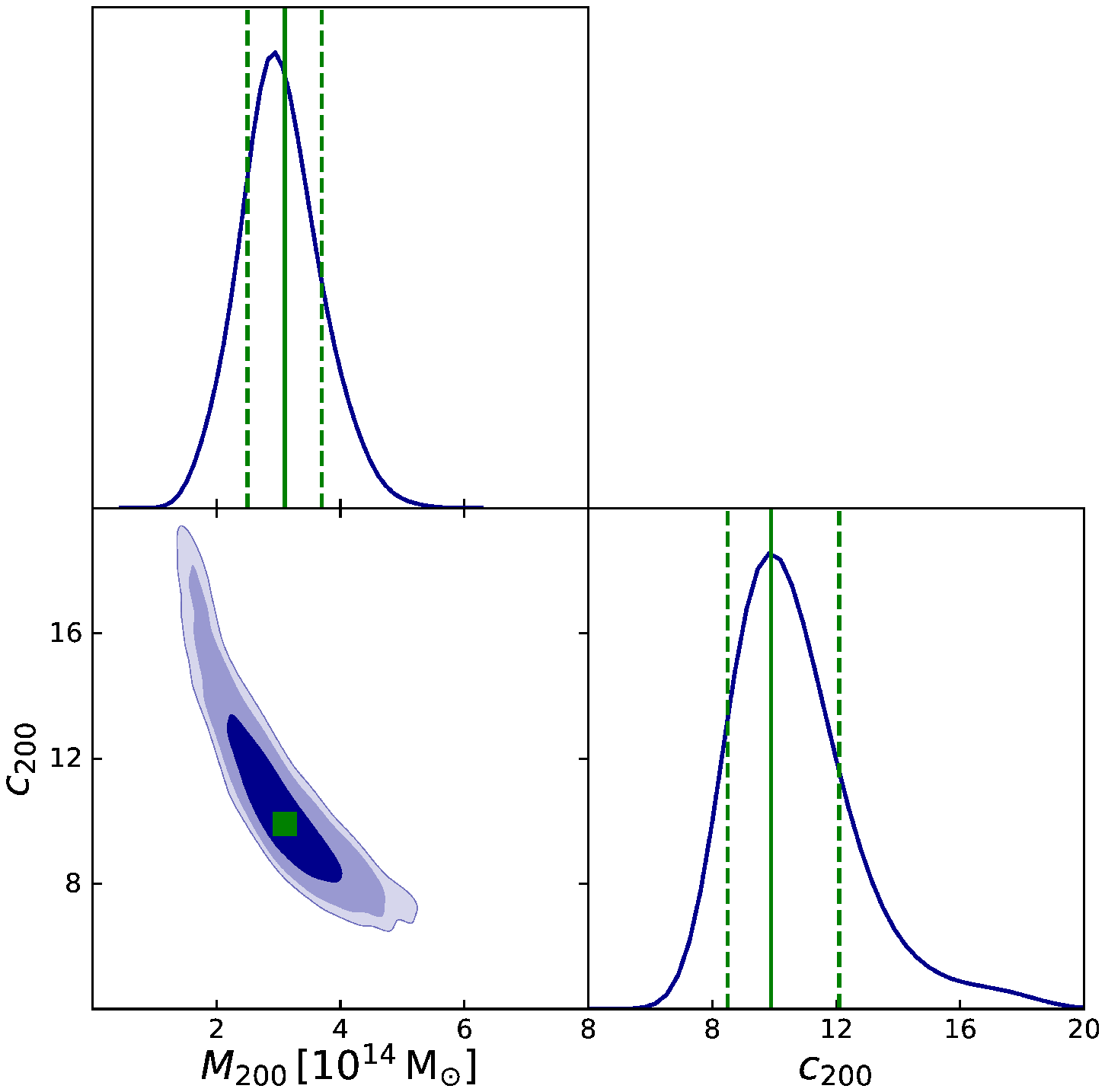}
\hspace{0.1cm}
\includegraphics[scale=0.35]{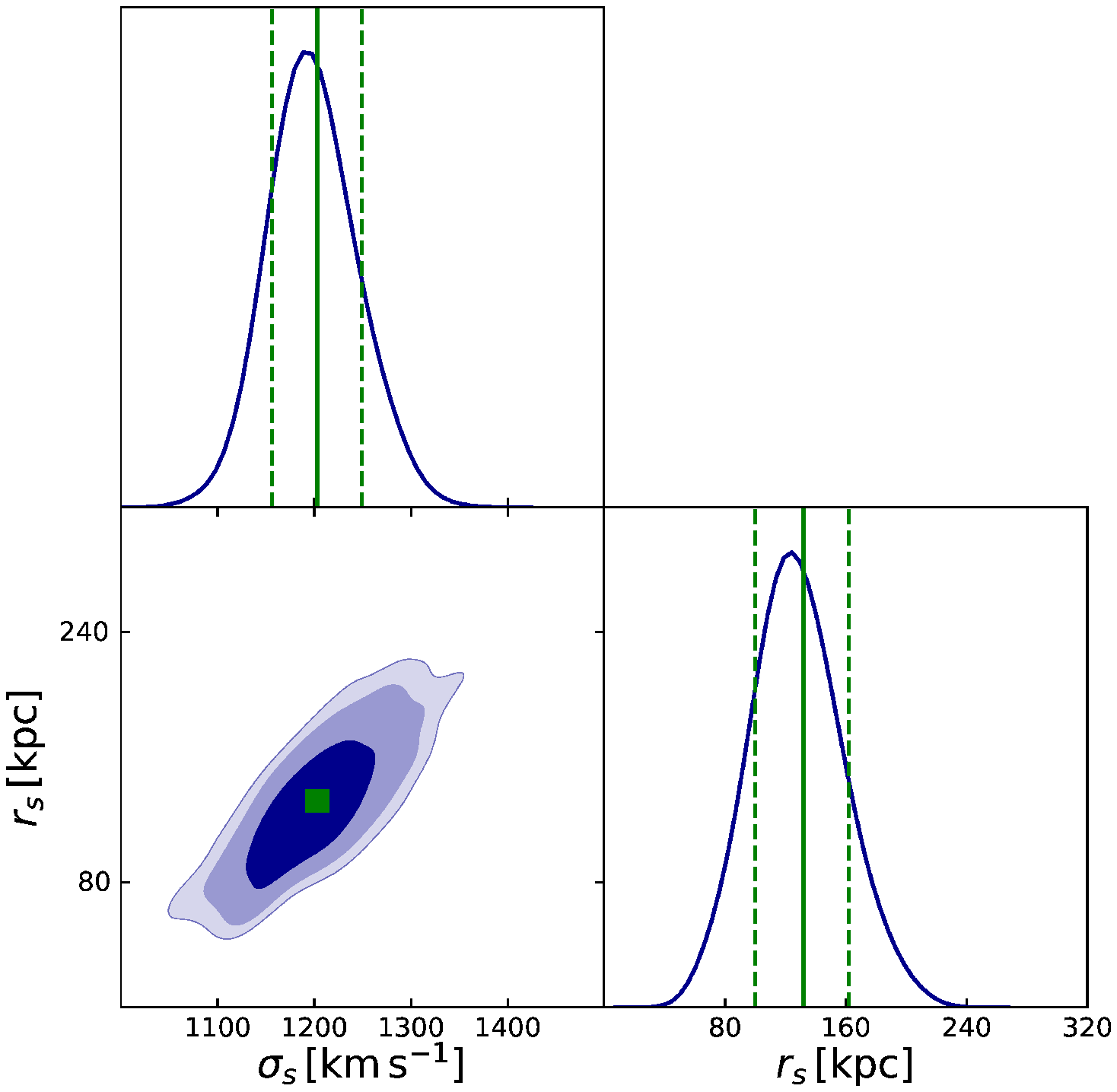}\\
\caption{\textit{Left column.} PDFs and  contours of the parameters $c_{200}$ and M$_{200}$. \textit{Right column.} PDFs and contours of the parameters $r_{s}$ and $\sigma_{s}$. From top to bottom \textrm{M$_{\textrm{lens}}$},  \textrm{M$_{\textrm{lens-$\sigma_s$}}$}, and \textrm{M$_{\textrm{lens-$\sigma_s$-mass}}$}, respectively. The three contours stand for the 68$\%$, 95$\%$, and 99$\%$ confidence levels. The values obtained for our best-fit model are marked by a green square, and by vertical lines in the 1D histograms (the asymmetric errors are presented in Table 2)} 

\label{SED_PDF}
\end{center}\end{figure*}

Considering the perturbation produced by galaxies 896(B) and 599(D) on different arclets (see Figure\,\ref{fig1}), some of their dark matter halo parameters were allowed to vary in the optimization procedure. Therefore, every model is computed and optimized in the image plane with a total of 17 free parameters that we detail below:  the six parameters, \{$X$,  $Y$, $\epsilon$,  $\theta$, $r_s$, $\sigma_{s}$\}, for the main halo,  two parameters,  \{$r_{\rm cut}^{*}$, $\sigma^*_0$\},  for the smaller-scale clumps,  the redshift, \{$z_{M5}$\}, for the arc system M5,   and finally, four parameters,  \{$\epsilon$,  $\theta$, $r_{\rm cut}$, $\sigma_0$\} for each of the galaxies 896(B), and 599(D). All the parameters are allowed to vary with uniform priors.  As an additional test of our best models (the three models presented in Section ~\ref{subsec:3models}), we used the fact that no counter image was identified for the spectroscopically confirmed background source, S1. Namely, with the parameters obtained with our best models  we use LENSTOOL to search for possible counter-images of S1, but in all the cases, S1 is returned as a single image. We try to include system M4 in our calculations with the redshift as a free parameter, but we were unsuccessful to reproduce the configuration. When we include this system the $\chi^{2}/_{DOF}$ is more than seven times the one obtained in our best models, i.e., $\chi^{2}/_{DOF}$  $\approx$ 15. Therefore, we exclude this system from our final models.  It is important to stress that we are not  claiming that M4 is not an arc system, just that the information is not enough to reproduce neither its configuration nor its redshift. Including this system would have increased the number of free parameters of the central galaxies, since their perturbation may affect the shape and position of the arcs.

\subsection{Weak Lensing} \label{WL}

To estimate the weak-lensing mass of MS0440.5+0204, we followed  the methodology described in detail in \cite{foex12}, which can be summarized as follows. The detection and selection of objects was performed in the r'-band with {\sc SExtractor} \citep{ber96} in dual mode. To sort stars, galaxies and false detections, we combined different criteria: the size of the objects compared with the  point spread function (PSF hereafter), the position in the magnitude/central flux diagram with respect to the {\it star branch}, and their stellarity index as given by the CLASS\_STAR parameter of {\sc SExtractor}. After this first step, we obtained a number density of $\sim23\,\mathrm{arcmin^{-2}}$ for galaxies and $\sim2\,\mathrm{arcmin^{-2}}$ for stars. To select the lensed sources, we removed galaxies located in the red sequence, estimated in the r'-(g'-r') diagram, down to a magnitude of $m_{r'}=23\,\mathrm{mag}$. All galaxies brighter than $m_{r'}=21\,\mathrm{mag}$ and fainter than $m_{r'}=25\,\mathrm{mag}$ were also removed. These cuts led to a number density of $\sim12\,\mathrm{arcmin^{-2}}$.

Next, we estimated the shape of the background galaxies with the software {\sc Im2shape} \citep{bridle02}, as presented in \cite{foex12,foex13}. Briefly, the catalogue of stars was used to estimate the local PSF, which was then interpolated at the position of each galaxy. The code convolves the PSF field to an elliptical model for the galaxy shape, and runs an MCMC sampler to find the parameters that minimize the residuals. The method was calibrated in the STEP1 simulations \citep{heymans06}, and the measured shear was corrected accordingly.

To estimate the lensing strength (i.e. the factor needed to convert shear into mass), we relied in the approach that consists in assuming that similar selection criteria applied on similar data sets must generate the same redshift distribution. We used the photometric redshifts from the T0004 release of the CFHTLS-DEEP survey that were computed with {\sc HyperZ} \citep{bolzonella00}.  After applying the same photometric selection criteria to the CFHTLS sample, we calculated for each galaxy the angular-diameter distances lens-source $D_{ls}$ and observer-source  $D_{os}$. By setting $D_{ls}$=0 whenever the photometric redshift is smaller than the cluster redshift, we automatically correct for the contamination of our sources catalogue by foreground galaxies. We estimate the average lensing strength to be $\beta=<D_{ls}/D_{os}>=0.66$, a  similar value to that of \citet{hoekstra2015} who used $\beta$=0.656 to derive their lensing mass estimate.

The shear profile was computed in logarithmically-spaced annuli, and fitted from 100 kpc to 2.5 Mpc. We used two parametric mass models: a singular isothermal sphere (SIS), and the classical NFW model. We propagated the uncertainties in the galaxy's shape parameters to the mass estimate by generating 5000 Monte Carlo realizations of the shear profile. Each of these profiles was obtained after randomly sampling the ellipticity parameters of each galaxy, assuming a normal distribution for the mean and the standard deviation given by Im2shape. The SIS and NFW best-fit parameters and their corresponding 1$\sigma$ uncertainties are given in Figure~\ref{fig2}, where we plot the average density contrast $\Delta\Sigma$ \cite[see][]{foex12,foex13}. This quantity is related to the shear and is calculated averaging the density contrast  within an annulus of projected physical radius $r$ around the center.\footnote{The density contrast of a circular-symmetric lens is estimated by the tangential shear $\gamma_{t,i}$ it produces on a galaxy source $i$ located at the concentric radius $r_i$ : $\Delta\Sigma$ = $\Sigma_{crit,i}$ $\times$ $\gamma_{t,i}$. Where the critical density reads, $\Sigma_{crit,i}$ = $c^2/4\pi G \beta D_{OL}$. And $D_{OL}$ is the angular-diameter distance observer-lens.}  The SIS and NFW models provide a good fit, with a reduced $\chi^2=9.1/8$ and $\chi^2=8.3/7$ for the SIS and NFW profiles, respectively. For the NFW density profile we obtain a mass $M_{200}$ = 3.9$^{+1.4}_{-0.8}$ $\times10^{14}\,$M$_{\odot}$, at R$_{200}$ = 1460$\pm$130 kpc. In  Figure~\ref{fig3} we show the two-dimensional contours on the NFW mass and concentration parameters, as estimated from the Monte Carlo realizations. Despite large error bars, we managed to constrain the concentration parameter, whose value $c_{200}\sim9$ is higher than expected for a cluster having a mass $M_{200}\sim4\times10^{14}\,M_{\odot}$ (e.g. \citealt{Duff08}). However, such a high concentration is in fairly good agreement with the $c(M)$ relation of \cite{foex14}, which was obtained by combining stacks of strong-lensing galaxy groups and clusters. We will return to this subject in the discussion section.

\textit{2D-shape\,-} To characterize the two-dimensional shape of the cluster from the weak-lensing data, we followed the same  approach as presented in \citet{Soucail2015}. It is based in the software {\sc LensEnt2} \citep{Marshall2002}, which uses the shape of each background galaxy as a local estimator of the reduced shear. It employs an entropy-regularized maximum-likelihood technique to produce a projected mass map of the cluster. The spatial resolution of the grid is chosen to have roughly one galaxy per pixel, leading to a similar number of constraints and free parameters. The mass map is the outcome of the convolution of the pixels grid (i.e. the free parameters of the model) with a broad kernel, to ensure a smoothed mass distribution. We chose a Gaussian kernel with a width of 100'', providing a good compromise between smoothness and details in the map. By sampling the probability distribution of each pixel, {\sc LensEnt2} produces an error map, whose average value around the cluster gives the expected level of uncertainty in the mass reconstruction. We used this value to produce the mass contours presented in Section ~\ref{sec:discussion}.\\

\begin{table}
\caption{FOM's}
\label{tbl-3} 
\centering 
\begin{tabular}{lccc}
\hline\hline 
\\
Model & $\chi_{DOF}^{2}$ &   \multicolumn{1}{c}{\textrm{FoM}/\textrm{FoM}$^{*}$}  &   \multicolumn{1}{c}{\textrm{FoM}/\textrm{FoM}$^{*}$}      \\
&   & [$r_s$-$\sigma_s$] &  [$c_{200}$-M$_{200}$]        \\
\\
\hline 
\\
\textrm{M$_{\textrm{lens}}$}   & $2.2$ & $0.58$ & $0.55$   \\
\textrm{M$_{\textrm{lens-$\sigma_s$}}$}  & $2.5$ & $0.93$ & $1.1\phantom{0}$   \\
\textrm{M$_{\textrm{lens-$\sigma_s$-mass}}$}  & $2.2$ & $1.0\phantom{0}$ & $1.0\phantom{0}$\\
\\
\hline 
\end{tabular}
\tablecomments{ \textrm{FoM}$^{*}$: \textrm{FoM} for model M$_{\textrm{lens-$\sigma_s$-mass}}$.}
\end{table}

\subsection{The three models} \label{subsec:3models}

\textit{Model \textrm{M$_{\textrm{lens}}$}\,-}  Our first strong lensing model uses only the  arc's positions as  constraints. Some of the arc systems show multiple subcomponents (bright knots) that can be conjugated as different multiple image systems \cite[e.g.,][]{bro05, Verdugo2011,Verdugo2016} increasing the number of constraints. Considering systems M1 (four images), M2 (decomposed in three systems with two images each), M3 (arranged in two systems with four images each), and M5 (two images), we get a total of 20 images (see Fig.~\ref{A00} in \ref{sec:ApendixA}). For these systems we have 26 constraints (if there is a total of $\sum_{i=1}^{n}n_i=N$ images, then there are $\sum_{i=1}^{n}2(n_i-1)=N_c$ constrains in the models assuming that the position, x,y of the images are fitted). 
This means 9 degrees of freedom for this model (17 free parameters). The results of our fits are summarized in Table~\ref{tbl-2}, where we present all de parameters of each model, in such a way that they can be used by the community.\footnote{The parameters files (from modeling)  containing the information presented in this section are available on request from the corresponding author.} The first column identifies the model parameters. The first two rows the position $X$, $Y$ in  arcseconds relative to the BCG. Rows 3$-$8  list the  parameters associated with the NFW profile: ellipticity, position angle, scale radius, velocity, concentration and mass. Rows (9)$-$(11) provide the parameters related to the galaxy scale clumps.  The results presented in this table allow us to compare parameters among the three different models. The Fig.~\ref{critics} shows the critical lines (for a source at $z$ = 0.95, $z$ = 1.10, and $z$ =  2.08) and the predicted positions of the images for systems M1, M2, M3, M5, and S1. We obtain $\chi^{2}/_{DOF}$ = 2.2 for this model (see Table~\ref{tbl-3}), with the most precise value for the redshift of system M5 equal to $z$ = 1.5$^{+0.2}_{-0.1}$.

In appendix (see Fig.~\ref{A1}) we present the 2D contours and the PDFs for the parameters of the main halo as well as $r_{cut}$ and $\sigma_0$ values for the galaxy scale clumps.  Although the degeneracy between parameters is common in lensing modeling, and has been studied previously \citep[see discussion in][]{jul07}, we want to show how these contours change when adding the additional constraints to the lensing models.  Note that the position of the large scale dark matter clump is found to be offset from the BCG.

\textit{Model \textrm{M$_{\textrm{lens-$\sigma_s$}}$}\,-}  Our second strong lensing model adds, as an additional constraint, the velocity dispersion of the cluster (see Eq.\,\ref{eq:Chi2Sigma}). This implies 10 DOF for this model. The value of the line-of-sight velocity dispersion was calculated in \citetalias{Rodrigo2020}, $\sigma_{obs}$ = 771$^{+63}_{-71}$ km s$^{-1}$, using 93 confirmed members. This velocity is slightly lower, but within the errors, than the value calculated from weak lensing analysis using an SIS, $\sigma_{WL}$ = 894$^{+52}_{-61}$ km s$^{-1}$ (see section\,\ref{WL}). In Table~\ref{tbl-2} we present the best-fit values, and  in the left panel of Fig.~\ref{A0} the critical lines and the predicted positions of the images. Comparing this figure with Fig.~\ref{critics}, we conclude that both models recover the image positions equally well, which is consistent with their similar  $\chi^{2}/_{DOF}$  (see also section \S \ref{sec:discussion}). The 2D contours and the PDFs for the free parameters are shown in Fig.~\ref{A2}.  We note that the model \textrm{M$_{\textrm{lens-$\sigma_s$}}$} produces similar critical lines and predicted
positions to those for the model \textrm{M$_{\textrm{lens}}$}.  We obtain $\chi^{2}/_{DOF}$ = 2.5 for this second model, with the best value for the redshift of system M5 equal to $z$ = 1.3$^{+0.2}_{-0.1}$.

\textit{Model \textrm{M$_{\textrm{lens-$\sigma_s$-mass}}$}\,-} Finally, we include the WL mass in our last model. From our weak lensing data we calculate the mass at radius $r_s$ and found $\tilde{m}_{WL}$  =  8.4 $\pm$ 2.5 $\times10^{13}\,M_{\odot}$. This value is used in Eq.\,\ref{eq:Chi2X} to set the final constrain to our model;  resulting in 11 DOF. In Table~\ref{tbl-2} we present the best-fit values,  and the right panel of Fig.~\ref{A0} exhibits the critical lines and the predicted positions of the images.  The 2D contours and the PDFs are shown in Fig.~\ref{A3}. For this model we obtain $\chi^{2}/_{DOF}$ = 2.2, with the best value for the redshift of system M5 equal to $z$ = 1.5$^{+0.2}_{-0.1}$.

\begin{figure*}[!htp]
\epsscale{1.1}
\plotone{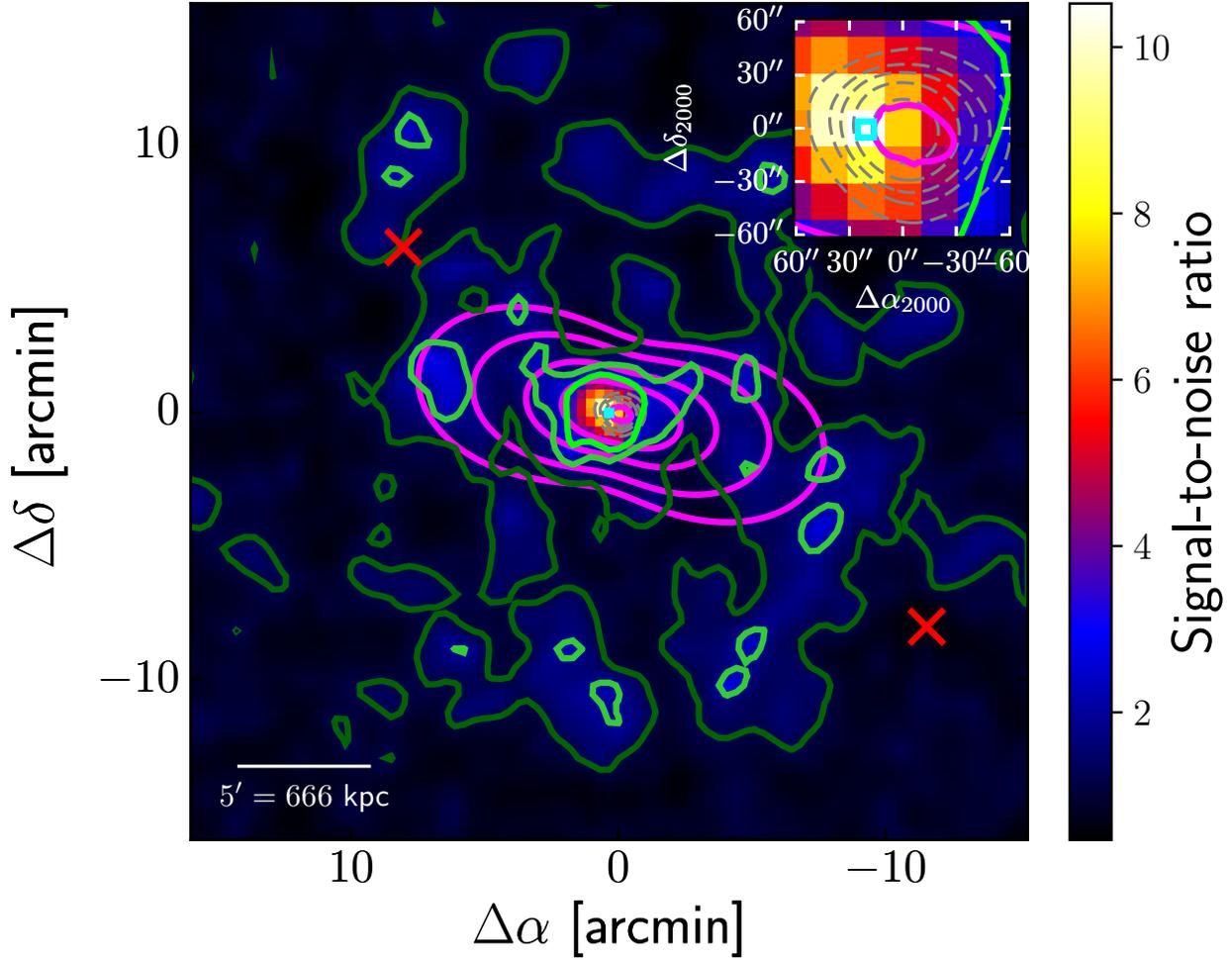}
\caption{ Weak lensing 2D mass map. The center of the map coincides with the galaxy 896(B) located at $\alpha$  = $\mbox{4:43:09.8}$, $\delta$ = $\mbox{+2:10:18.2}$.  The three different shades of green color represent signal-to-noise ratios of  3, 2, and 1.2 respectively. The magenta-contours correspond to the strong lensing mass contours.  The cyan square show the position of the peak of the lensing mass map. The two red crosses represent the position of the clusters  ZwCl0441.1+0211, and CLJ0442+0202, top-left and bottom-right, respectively. The surface brightness from Chandra observations (gray dashed contours) is also shown in the inset for clarity.\label{WLSLMapp}}
\end{figure*}

\section{Discussion}\label{sec:discussion}

\subsection{Including additional constraints}

The degeneracies depicted in Figures ~\ref{A1}, ~\ref{A2}, and ~\ref{A3} are expected;  and similar results  have been presented previously by different authors \citep[e.g.,][]{jul07,eli07}.  We can appreciate that the 2D contours and  the best values for $\sigma_0^{*}$ and $r_{cut}^{*}$  change significantly in each model, reflecting a degeneracy in mass between the smooth cluster component and the galaxies. Another interesting degeneracy is the one between  $\sigma_s$ and the scale radius, $r_s$, which is related to the definition of the gravitational potential \citep[see][]{jul07}.  The right panel of Fig.~\ref{SED_PDF}  shows the PDFs and the contours of these parameters for the three models discussed in this work. Note that the inclusion of dynamical constraints reduces the 2D contours (compare top and middle right panels of Fig.~\ref{SED_PDF}), which reflects the fact that the parameters are related through Eq\,~\ref{eq:sigmaNFW_pro}. Since the parameter $\sigma_s$ is constrained by the measured velocity dispersion of the cluster (see Eq.\,\ref{eq:Chi2Sigma}), the constrain is spread to the scale radius parameter $r_s$. The tightening of the contours also decreases the errors associated with the parameters. For example, for $r_s$ the relative error between \textrm{M$_{\textrm{lens}}$} and \textrm{M$_{\textrm{lens-$\sigma_s$}}$} changes from 0.30$\%$ to 0.25$\%$.

The left panel of Fig.~\ref{SED_PDF} also presents the PDFs and the contours for $c_{200}$ and M$_{200}$. 
We observe an analogous behavior: a reduction of the contours when adding additional constraints.  As discussed in \citet{Verdugo2016}, this result is expected since dynamics do constrain the scale radius of the NFW mass profile, a parameter that is not accesible to strong lensing alone.  Note also (Fig.~\ref{SED_PDF}) that the inclusion of the mass as an additional constrain in model \textrm{M$_{\textrm{lens-$\sigma_s$-mass}}$}   improves, although slightly, the result  and  tightens the 1D histograms (compare middle and bottom panels). In general the contours in the  \textrm{M$_{\textrm{lens-$\sigma_s$-mass}}$} are better defined, with the best solution nearly in the center of the 1$\sigma$ region.  This suggests that the inclusion of the WL mass is complementary to the use of velocity dispersion. This is interesting since WL (as well as SL) provides the total amount of mass and its distribution with no assumptions about the dynamical state of the matter producing the lensing effect.

Interestingly, the three models are equally good to reproduce the image positions of the arcs, which is quantified not only through the $\chi^{2}/_{DOF}$ (very similar in the models) but also through their \textit{rmsi} (root-mean-square image)\footnote{The root-mean-square image is defined through the expression: $\textit{rmsi} =   \sqrt{ \frac{1}{n}  \sum_{j=1}^{n_i} \left[ x_{obs}^j - x^j(\theta)    \right]^2 }$. Where $n$ is the number of images for the system. }.  This value provides another way to quantify the goodness of the model \citep[see][]{eli07,lim07b}. We obtain \textit{rmsi} = 0.50, 0.56, and 0.54 for \textrm{M$_{\textrm{lens}}$} , \textrm{M$_{\textrm{lens-$\sigma_s$}}$}, and \textrm{M$_{\textrm{lens-$\sigma_s$-mass}}$}, respectively.  Their similar values are consistent with the predicted positions of the lensed images, and the critical lines depicted in Figures ~\ref{critics}, and ~\ref{A0}; they look extremely similar. We can also appreciate that our models do not predict any counter-images for the single image S1, in accordance with our assumption that this is a distorted image of a single source.

Finally, as a supplementary comparison between models, we calculate the inverse of the square root of the the covariance matrix. This figure-of-merit 
\citep[so-called \textrm{FoM}, see][and also  \citealp{Magana2015} for an application to lensing models]{Albrecht2006}  quantifies the ability of the observational data set to constrain the parameters; a larger \textrm{FoM} means stronger constraints. For the parameters $r_s$-$\sigma_s$, the  ratios between \textrm{FoM's}, relative to model \textrm{M$_{\textrm{lens-$\sigma_s$-mass}}$} are 0.58, 0.93, and 1 respectively (see Table~\ref{tbl-3}).  This confirms that the inclusion of dynamical restrictions (\textrm{M$_{\textrm{lens-$\sigma_s$}}$}) gives slightly more stringent constraints than those provided by the model \textrm{M$_{\textrm{lens}}$}. The  incorporation of the mass in model \textrm{M$_{\textrm{lens-$\sigma_s$-mass}}$}  increases the \textrm{FoM}, however not appreciably. Similarly, we observe the same behavior for the parameters  $c_{200}$-M$_{200}$.  We conclude that the three models are equally good in reproducing the strong lensing features presented in \object{MS\,0440.5+0204}. However, they are different at large scale, with a slightly better performance for model \textrm{M$_{\textrm{lens-$\sigma_s$-mass}}$}. This is because the strong lensing model is sensitive to the mass distribution at inner radii, whereas the dynamics \citep{Verdugo2016} as well as the WL mass provide constraints at larger radius. Thus, in the next sections we concentrate our discussion of the \textrm{M$_{\textrm{lens-$\sigma_s$-mass}}$} model as the favored model.

\subsection{Over-concentration}

Galaxy clusters with prominent strong lensing features commonly have high concentration values when modeled with NFW profiles \citep[e.g.,][]{gav03,kne03,bro05,Comerford2007,Oguri2009}, larger than those predicted by N-body simulations of cluster formation and evolution in a $\Omega_{\Lambda}$-dominated Universe \citep[e.g.,][]{bul01}. This over-concentration is interpreted as the cluster dark matter halo departure from spherical symmetry \citep[e.g.,][]{gav05},  with the major axis of their triaxial geometry oriented towards the line-of-sight \citep[e.g.,][]{Corless2007,Corless2009}. Indeed, strong-lensing clusters could represent a biased population with their major axes aligned with the line-of-sight \citep{Hennawi2007,OguriBlandford2009,Meneghetti2010,Giocoli2016}. Thus, a value of $9.9^{+2.2}_{-1.4}$ (Table~\ref{tbl-2}) for a galaxy cluster as \object{MS\,0440.5+0204} it is not surprising. Besides that, this value is consistent with the value obtained from weak lensing, $c_{200}$ = 9.3$^{+4.8}_{-3.5}$, which supports the idea of a highly concentrated cluster. In the left panel of Fig.~\ref{SED_PDF} we show the PDFs and the contours for the parameters $c_{200}$ and $M_{200}$, which display similar shape among models. This high concentration is consistent with the analysis presented in  \citetalias{Rodrigo2020}, since the cluster  displays evidence of having an elongation along the line of sight. Interestingly, the redshift distribution of the galaxies reveals three sub-clumps (see \citetalias{Rodrigo2020}), indicating that the cluster core might be experimenting a merging process along the line-of-sight. This evidence of substructures is important since, as mentioned before, they can bias the strong lensing results \citep{Bayliss2014}, as they are sensitive to the projected mass along the line of sight.

\subsection{Is MS\,0440.5+0204 part of a  larger structure?}

We discuss here the projected mass map from the best-fit model of \textrm{M$_{\textrm{lens-$\sigma_s$-mass}}$}.  Given our constraints, the strong lensing mass is reliable up to the scale radius ($r_s$ = 132$_{-32}^{+30}$), a value beyond which the mass is extrapolated. However, as we will see later, this is consistent with other measurements. Figure~\ref{WLSLMapp} shows the 2D mass map from the strong lensing model (magenta lines) superimposed onto the 2D mass map from weak lensing. The strong lensing contours show the projected surface mass density at values of   3, 5, 10, 20, 150  $\times10^{8}$\,M$_{\odot}$ arcsec$^{-2}$.  Interestingly, their direction (position angle) is consistent with the one from the contours of the 2D weak lensing signal. Furthermore,  we use the task \textit{ellipse} in IRAF to fit ellipses to the smoothed X-ray image of the cluster to obtain the position angle. We find a value of 160$^{\circ}$ $\pm$ 4$^{\circ}$, in agreement with the position angle obtained from strong lensing (see Table~\ref{tbl-2}). The elongation in the \object{MS\,0440.5+0204} mass is likely produced by the influence of a second structure (a group or galaxy cluster) in the North-East direction. The red cross in the top-left corner of Figure~\ref{WLSLMapp} depict the position of the cluster \object{ZwCl0441.1+0211}, unfortunately this object does not have a reported redshift (see discussion in \citetalias{Rodrigo2020}).

Note also that the peak of the weak lensing mass contours  (dark-green square on Figure~\ref{WLSLMapp}) is shifted with respect to the center of the cluster (i.e. the BCG), which is tempting to associate with the presence of another galaxy cluster. However, the uncertainty in the position of the mass peak has a lower limit of 29$\arcsec$ ($\approx$ 94 kpc), which means that the position of the BCG is within the measurement error. Analyzing 25 galaxy clusters, \citet{Oguri2010} found that the mass centroid obtained from weak lensing can be constrained with an accuracy of 50 kpc in radius, but some clusters showed significant offsets, up to 100 kpc. These kind of offsets are common, and could be caused by shape noise \citep{Dietrich2012}. 
\citet{Martinet2016} showed that fake lensing peaks are expected in weak lensing maps due to sampling and shape noise.
This effect could be responsible for the weak lensing peak near the position of \object{ZwCl0441.1+0211}.

\begin{figure*}[!htp]
\epsscale{0.6}
\plotone{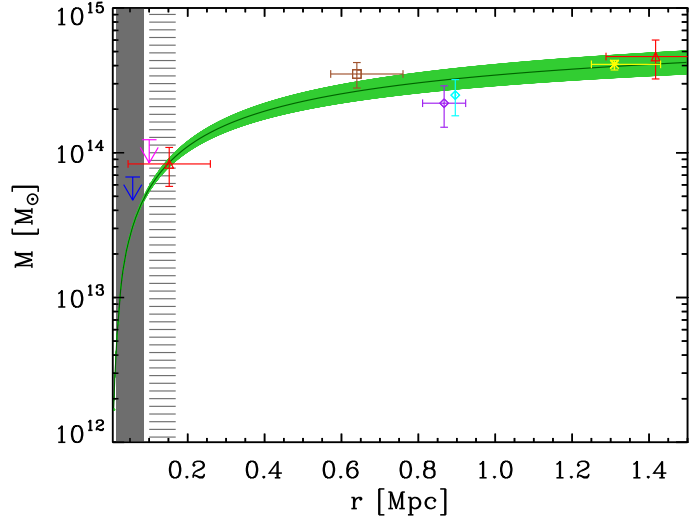}
\caption{2D projected mass as a function of the radius measured from galaxy 896(B). The dark green-line, and green-shaded area correspond to the mass profile and its respective 1-$\sigma$ error for the strong lensing model. The grey-shaded region shows the area where the arc systems lie. The region filled with horizontal grey-lines depict the best $r_s$ value from strong lensing. The yellow-asterisk  marks the projected mass calculated from dynamics \citepalias{Rodrigo2020}.  The brown open-square  indicates the projected X-ray mass calculated at radius $R_{2500}$ from the data reported in \citet{Hicks2006}. The two red triangles designate our two estimates (at $r_s$ and $r_{200}$) from the weak lensing mass. The purple and the cyan-diamond  denote the weak lensing mass at radius $r_{500}$ from \citet{hoekstra2012}, and \citet{hoekstra2015}, respectively. We also depict with  blue and magenta arrows the upper limits for the mass from \citet{Gioia:1998} and  \citet{Shan2010}, respectively.\label{fig5}}
\end{figure*}

\subsection{Comparison with other works}\label{sec:comparison}

The projected mass as a function of radius obtained by integrating the  2D dimensional map (for model \textrm{M$_{\textrm{lens-$\sigma_s$-mass}}$}) is shown in Fig.~\ref{fig5}. The grey-shaded region shows the area where the arc systems lie, i.e.  from the nearest radial arc to the furthest tangential one from the center. Assuming a flattened potential \citep{bla87}, \citetalias{Gioia:1998} derived a projected mass distribution of $6.6-9.5 \times10^{13}$ M$_{\sun}$ for the central 24$\arcsec$ region encircled by the arcs. Their model was constructed assuming that the multiple images systems were the product of five different background sources (see Table~\ref{tbl-1}). Our results are consistent with this upper value (see blue-arrow in Fig.~\ref{fig5}).

 \object{MS\,0440.5+0204} was part of the sample of  \citet{Sand2005} who studied gravitationally lensed arcs in clusters of galaxies. The authors reported the length-to-width ratio of the arcs, and their magnitudes, but not the mass. \citet{Shan2010}  calculated the mass from strong lensing, but assuming a spherical matter distribution, and using $m_{lens}$ = $\pi r^2_{arc} \Sigma_{crit}$. As the authors point out, their model is not realistic and probably overestimates the mass. They reported a value $m_{lens}$ =  1.23 $\times10^{14}\,M_{\odot}$ within a radius of $r_{arc}$ = 0.1 Mpc. From our best model, we find a projected  mass of 5.6$\pm$0.3 $\times10^{13}\,M_{\odot}$ inside the same radius, which could explain the discrepancy between the $m_{lens}$ and $m_{Xray}$ reported by these authors ($m_{lens}$/$m_{Xray}$ $\sim$ 3). Indeed, using the values of $kT$, $r_c$, and $\beta$ reported by \citet{Hicks2006}, we calculated the projected mass inside the radius $R_{2500}$ following the procedure of \citet{Wu1994,wu00}. We found M$_{Xray}$ = 3.5 $\pm$ 0.7 $\times10^{14}\,M_{\odot}$, which  agrees, within the errors, with our mass estimate for the same radius (see Fig.~\ref{fig5}). In the same figure we note also that, although the values for the  mass at $r_{500}$ reported by \citet{hoekstra2012} and \citet{hoekstra2015} are slightly  smaller than the one obtained from our lensing model,  they agree within the errors.  \citet{hoekstra2015} performed a weak-lensing analysis of  \object{MS\,0440.5+0204} using the same dataset but with a different approach. In particular, they did not directly fit the concentration parameter of the NFW model but rather relied on the mass-concentration [c(M)] relation from \citet{Dutton2014}. Our results suggest that the concentration of \object{MS\,0440.5+0204} is higher than expected from these c(M) relations, therefore we can anticipate different results in the inferred weak-lensing mass estimate. \citet{hoekstra2015} quote a mass $M_{500}$ = 3.0$\pm$1.5 $\times10^{14}\,$M$_{\odot}$  at a radius $R_{500}$=1.28 Mpc. Within the same aperture, our best-fit model gives a mass $M$(1.28 Mpc)=3.5$\pm$0.9 $\times10^{14}\,$M$_{\odot}$, which is $\approx$15$\%$ larger than the results reported by \citet{hoekstra2015} but in agreement within the uncertainties.

\section{Conclusions}\label{sec:conclusions}

In  this work we have reconstructed the 2D mass distribution of \object{MS\,0440.5+0204}. We have presented a detailed strong lensing analysis of the core of this galaxy cluster, including four multiple imaged systems as constraints,  three of them with spectroscopic confirmation. 
Extending the investigation at large cluster centric distance, we have combined the strong lensing
analysis with other data sets calculated independently, namely the velocity dispersion and the weak lensing mass.  
Our analysis is threefold: a strong lensing only mass model (in which 13 galaxies have been added as perturbation to the cluster potential);
a model combining strong lensing with dynamical information (following the method of \citet{Verdugo2011}); a model combining strong lensing,
dynamics and the weak lensing estimated mass.

\citet{Verdugo2016} have demonstrated that it is possible to determine both the scale radius and the concentration parameter when combining lensing and dynamical constraints. In the present work we have reached a similar result.  Including additional  information (from velocity dispersion and mass) allows us  to reduce the degeneracy between  $\sigma_s$, and the scale radius $r_s$, or the mass and the concentration. However, those constraints do not improve the model (regarding the $\chi^{2}$, and the \textit{rmsi}) in the sense that the new data sets do not considerably affect the predicted positions of the lensing images. Although \textit{rmsi} values smaller than 1$\arcsec$ as the ones found in this paper (0.50$\arcsec$, 0.56$\arcsec$, and 0.54$\arcsec$) have also been reported in well modeled clusters \citep[e.g.,][]{new13,Jauzac2015,Kawamata2016}, they remain unaffected even  after imposing tight constraints in the velocity $\sigma_s$, and the scale radius.

\object{MS\,0440.5+0204} seems to be a highly concentrated galaxy cluster (either calculated with weak lensing $c_{200}$ = 9.3$^{+4.8}_{-3.5}$, or with strong lensing alone $c_{200}$ = 8.3$^{+3.2}_{-1.0}$), suggesting  that the major axis is probably oriented along the line of sight. This result is supported by the analysis presented in \citetalias{Rodrigo2020}. Although it is tempting to associate \object{MS\,0440.5+0204} with \object{ZwCl0441.1+0211}, our weak lensing analysis is not conclusive. More spectroscopic data is necessary to shed light on this problem.

\acknowledgments

The authors thank the anonymous referee for a careful reading of the submitted paper, and the thoughtful comments that helped to improve the document. This research has been carried out thanks to PROGRAMA UNAM-DGAPA-PAPIIT IA102517. ML acknowledges CNRS and CNES for support. J.M. acknowledges the support from CONICYT/FONDECYT project 316064. V.M. acknowledges support from Centro de Astrof\'isica de Valpara\'iso. JAD is grateful for the support from the UNAM-DGAPA-PASPA 2019 program and the kind hospitality of the IAC. We thank E. Ellingson for sharing her data.

\facilities{Gemini South: GMOS-S; CFHT: Megaprime/Megacam}

\newpage
\begin{appendix}

\section{Substructures in arc systems}\label{sec:ApendixA}

\begin{figure}[h!]\begin{center}
\includegraphics[scale=1.]{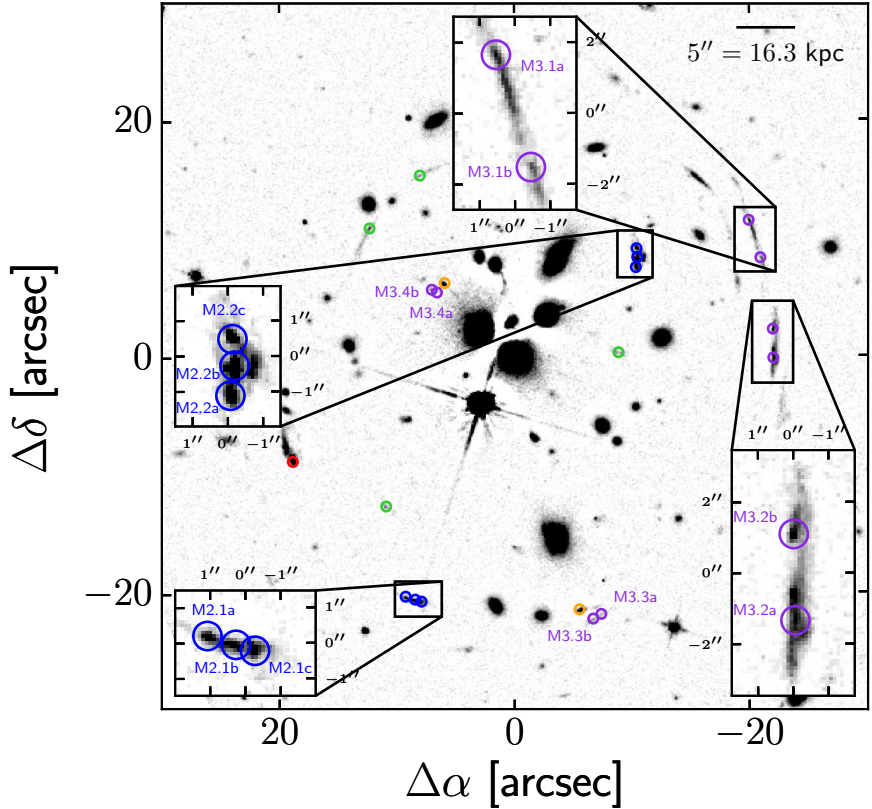}
\caption{HST WFPC2 F702W image, with local median average subtracted of the central region ($60\arcsec \times60\arcsec$) of the MS0440 cluster. The circles show the positions of the images (input data for the model). We follow Fig.1 code-colors for each family of lensed images.   \label{A00}}
\end{center} 
\end{figure}

Fig.~\ref{A00} show the position of the images used to construct the models presented in our work. To emphasize the substructure in some of the arcs systems, we zoomed in arc M2.1, M2.2, M3.1, and M3.2.  Note the multiple knots in system M2.

\section{Results for models M$_{\textrm{lens-$\sigma_s$}}$ and M$_{\textrm{lens-$\sigma_s-mass$}}$}

\begin{figure}[h!]\begin{center}
\includegraphics[scale=0.62]{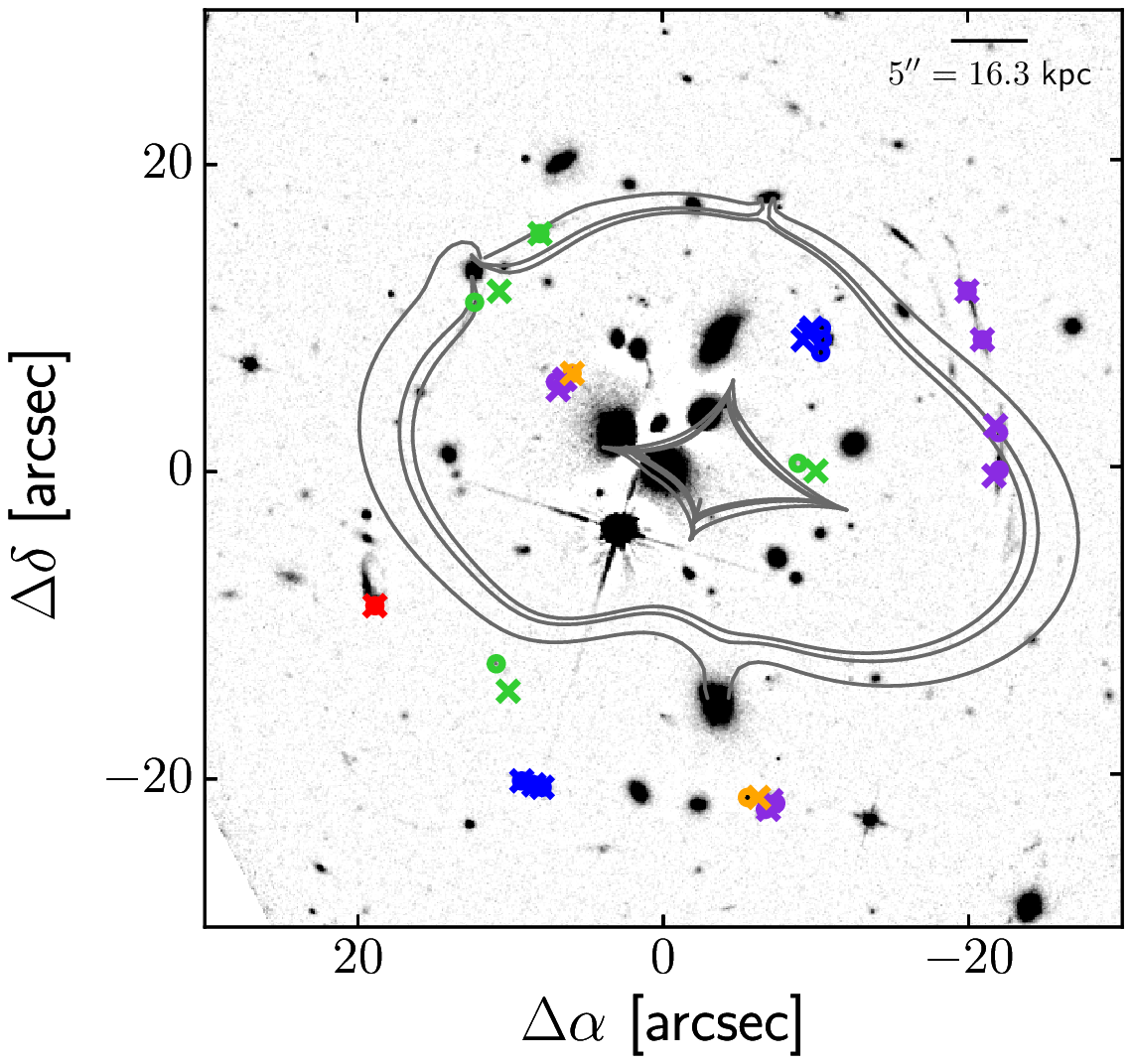}
\includegraphics[scale=0.62]{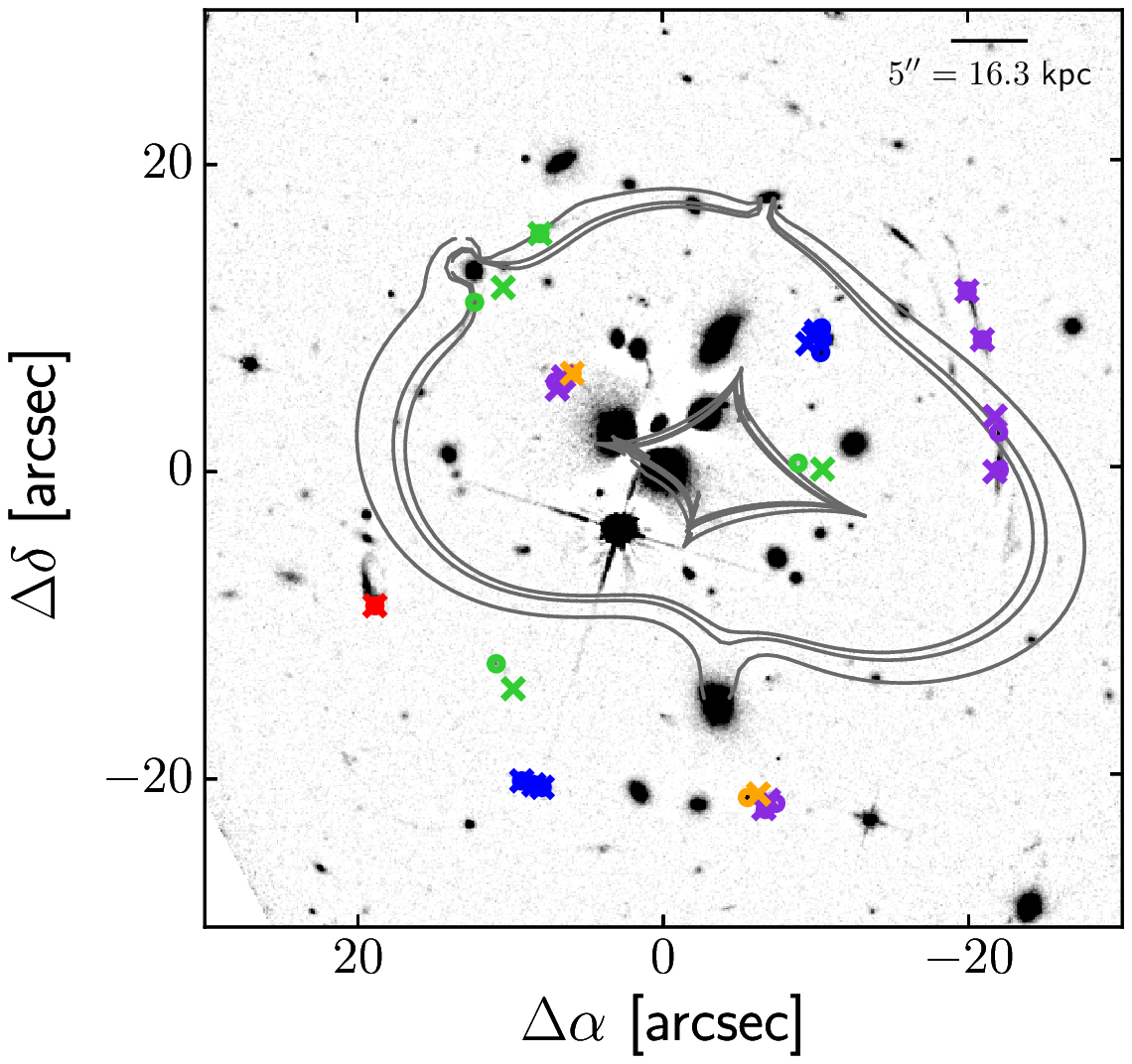}
\caption{Critical and caustic lines for a source located at $z$ = 0.95, $z$ = 1.10, and $z$ =  2.08, from inner to outer radii respectively. The circles show the positions of the images (input data for the model), and the crosses the predicted positions of the lensed images. Left panel .- Results from model M$_{\textrm{lens-$\sigma_s$}}$. Right panel .- Results from model M$_{\textrm{lens-$\sigma_s-mass$}}$.  We follow Fig.1 code-colors for each family of lensed images.  \label{A0}}
\end{center} 
\end{figure}

In order to compare the results of model \textrm{M$_{\textrm{lens}}$} (see  Fig.~\ref{critics}) with the two other models, i.e., M$_{\textrm{lens-$\sigma_s$}}$ and M$_{\textrm{lens-$\sigma_s-mass$}}$, the Fig.~\ref{A0} shows the predicted positions of the lensed images. Note how the three models recover the image positions equally well, and the critical and caustic lines are very similar.

\newpage

\section{PDFs and 2D PDFs of the model parameters.}

The next figures show the PDFs for the three models discussed in this work. The parameters exhibit the same degeneracies in all models since those depend on the lensing configuration  \citep[see][]{jul07}. However, incorporating additional constraints to the lensing models, namely the velocity and the mass, yield better constrained parameters with reduced contours.

 \begin{figure*}[!htp]
\centering
\includegraphics[scale=0.48]{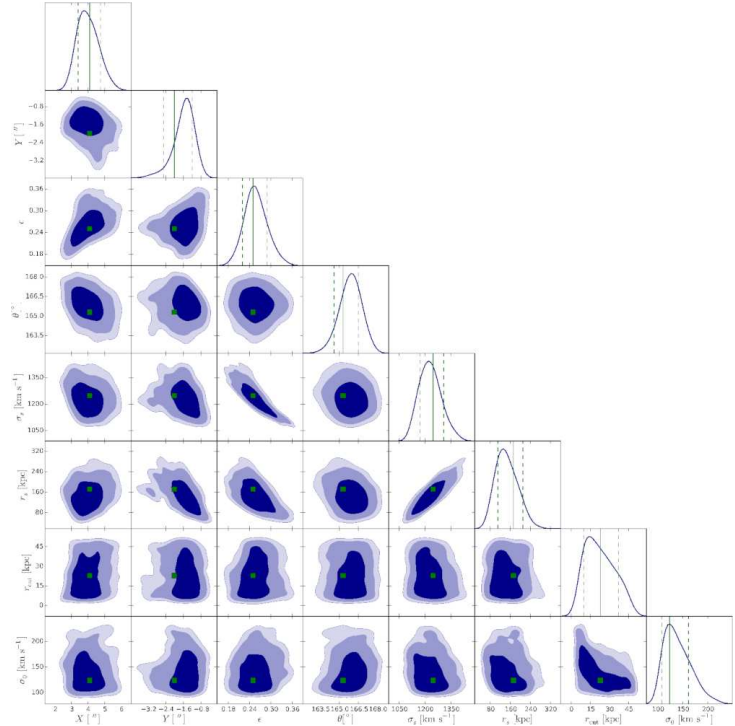}
\caption{PDFs and contours of the model parameters for \textrm{M$_{\textrm{lens}}$}. The three contours stand for the 68\%, 95\%, and 99\% confidence levels. The values obtained for our best-fit model are marked by a green square, and with vertical lines in the 1D histograms (the asymmetric errors are presented in Table~\ref{tbl-2}).}
\label{A1}
\end{figure*}

\clearpage

 \begin{figure*}[!htp]
\centering
\includegraphics[scale=0.48]{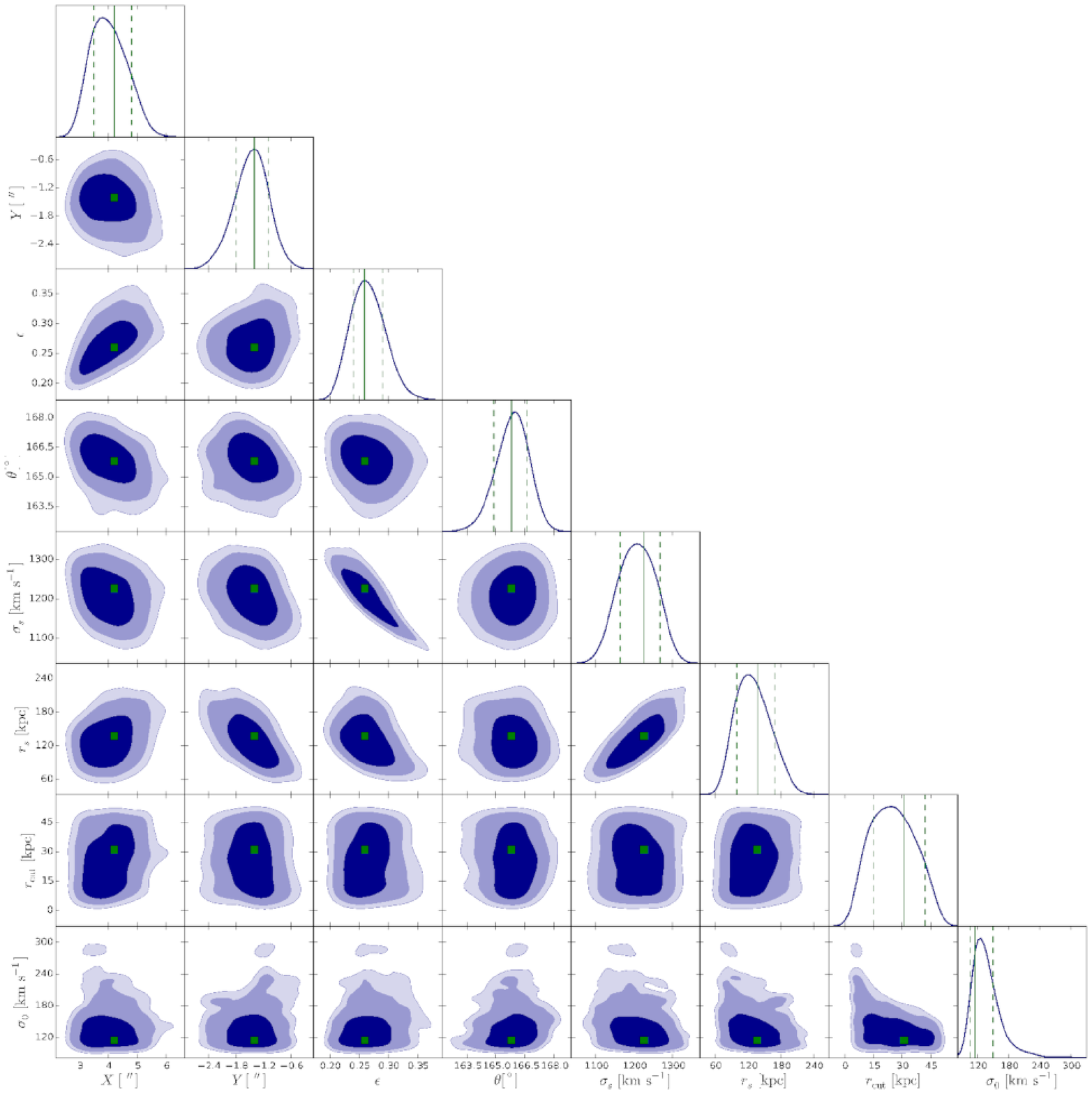}
\caption{PDFs and contours of the model parameters for \textrm{M$_{\textrm{lens-$\sigma_s$}}$}. The three contours stand for the 68\%, 95\%, and 99\% confidence levels. The values obtained for our best-fit model are marked by a green square, and with vertical lines in the 1D histograms (the asymmetric errors are presented in Table~\ref{tbl-2}).}
\label{A2}
\end{figure*}

\clearpage

 \begin{figure*}[!htp]
\centering
\includegraphics[scale=0.42]{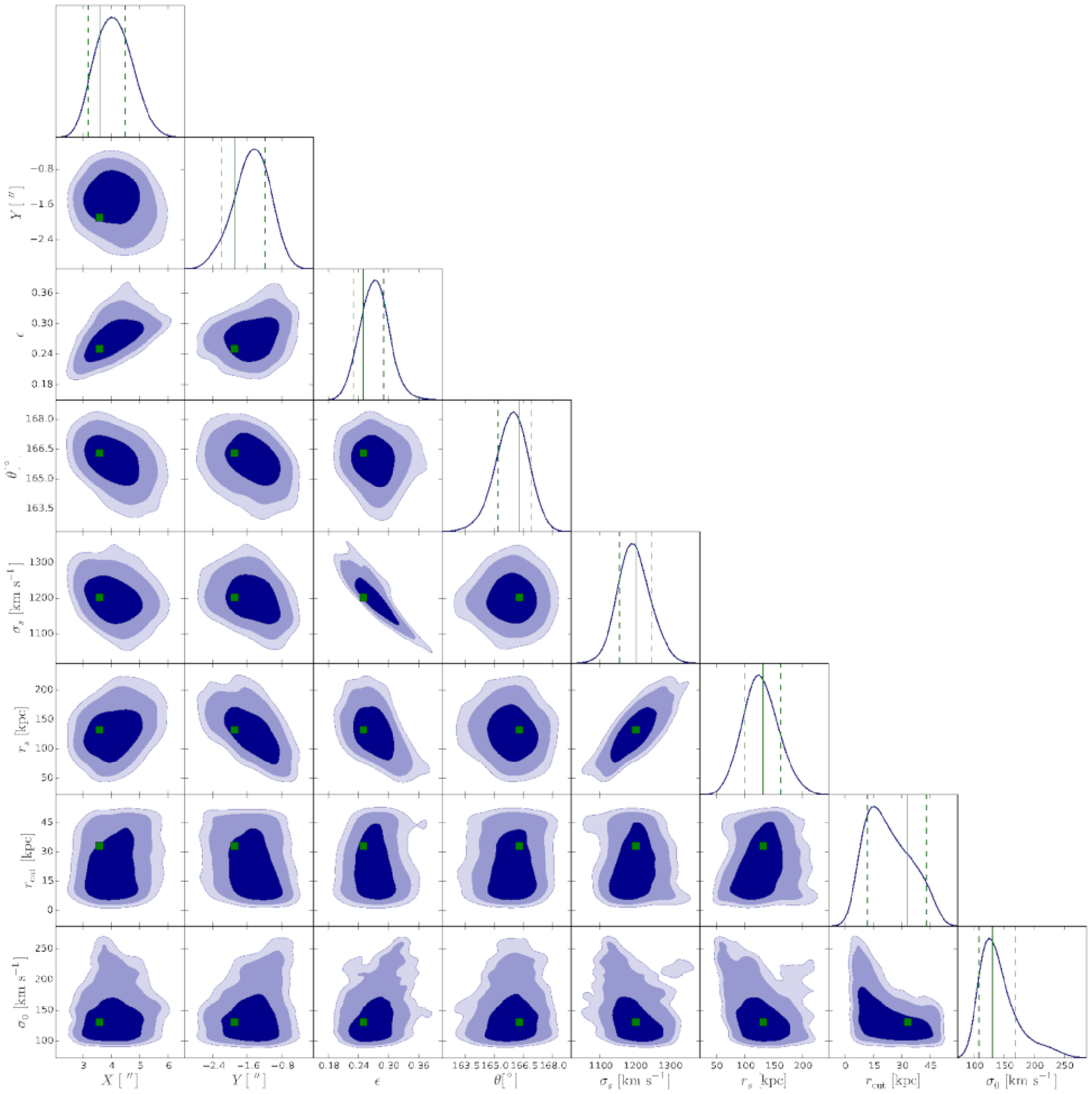}
\caption{PDFs and contours of the model parameters for \textrm{M$_{\textrm{lens-$\sigma_s$-mass}}$}. The three contours stand for the 68\%, 95\%, and 99\% confidence levels. The values obtained for our best-fit model are marked by a green square, and with vertical lines in the 1D histograms (the asymmetric errors are presented in Table~\ref{tbl-2}).}
\label{A3}
\end{figure*}

\end{appendix}

\clearpage

\bibliography{references}
\bibliographystyle{aasjournal}

\end{document}